\documentclass[journal]{IEEEtran}
\usepackage{graphicx}
\usepackage{caption}

\usepackage{multicol}
\usepackage{amsmath,amssymb}
\usepackage{amsfonts}
\usepackage{textcomp}
\usepackage{psfrag}
\usepackage{multimedia}
\usepackage{fancybox}
\newcommand{\vect}[1]{\ensuremath{\boldsymbol{\mathrm{#1}}}}

\newtheorem{Proposition}{Proposition}

\newtheorem{Lemma}{Lemma}

\usepackage[usenames,dvipsnames]{color}
\definecolor{wheat}{rgb}{0.96,0.87,0.70}
\definecolor{mario}{rgb}{0.8,0.8,1}
\definecolor{seb}{rgb}{0.8,1,0.8}

\newcommand {\matr}[2]{\left[\begin{array}{#1}#2\end{array}\right]}

\newcounter{lastnote}

\usepackage[ruled,vlined]{algorithm2e}
\begin{document} 

\title{Towards Safe Reinforcement Learning Using NMPC and Policy Gradients: Part I - Stochastic case} 

\author
{S\'ebastien Gros, Mario Zanon
	\thanks{S\'ebastien Gros is with the Department of Cybernetic, NTNU, Norway.}
	\thanks{Mario Zanon is with the IMT School for Advanced Studies Lucca, Lucca
		55100, Italy.}
}

%\authorrunning{S. Gros and Al.}%, D. Schlipf}
%\titlerunning{$\mathrm{SO}(3)$ Constraints in Periodic Optimal Control}
% Include the date command, but leave its argument blank.

%\date{}

%\title{$\mathrm{SO}(3)$ Constraints in Periodic Optimal Control}
%%\maketitle
%%\thispagestyle{empty}
%%\pagestyle{empty}
%
%
%\author{\IEEEauthorblockN{S\'ebastien Gros\IEEEauthorrefmark{1},
%		Mario Zanon\IEEEauthorrefmark{1}}
%	\IEEEauthorblockA{\IEEEauthorrefmark{1}Dept. Signals and Systems, Chalmers University of Technology, G\"oteborg, Sweden}% <-this % stops an unwanted space
%	\thanks{
%		Corresponding author:  (email: ).}}

%\textcolor{red}{
%\begin{itemize}
%\item Need to clarify the relationship "continuous-discrete", as an asymptotic relationship unless we can establish something about discretization schemes...
%\item should we also discuss SOSC ? It would be nice, but make the paper much longer...
%\end{itemize}
%}
\IEEEtitleabstractindextext{
	\begin{abstract} We present a methodology to deploy the stochastic policy gradient method, using actor-critic techniques, when the optimal policy is approximated using a parametric optimization problem, allowing one to enforce safety via hard constraints. For continuous input spaces, imposing safety restrictions on the stochastic policy can make the sampling and evaluation of its density difficult. This paper proposes a computationally effective approach to solve that issue. We will focus on policy approximations based on robust Nonlinear Model Predictive Control (NMPC), where safety can be treated explicitly. For the sake of brevity, we will detail safe policies in the robust linear MPC context only. The extension to the nonlinear case is possible but more complex. We will additionally present a technique to maintain the system safety throughout the learning process in the context of robust linear MPC. This paper has a companion paper treating the deterministic policy gradient case.
	\end{abstract}
	
}

\maketitle

\IEEEdisplaynontitleabstractindextext

\IEEEpeerreviewmaketitle

\section{Introduction}
Reinforcement Learning (RL) is a powerful tool for tackling Markov Decision Processes (MDP) without depending on a detailed model of the probability distributions underlying the state transitions. Indeed, most RL methods rely purely on observed state transitions, and realizations of the stage cost $L(\vect s,\vect a)\in\mathbb{R}$ assigning a performance to each state-input pair $\vect s,\vect a$ (the inputs are often labelled actions in the RL community). RL methods seek to increase the closed-loop performance of the control policy deployed on the MDP as observations are collected. RL has drawn an increasingly large attention thanks to its accomplishments, such as, e.g., making it possible for robots to learn to walk or fly without supervision \cite{Wang2012b,Abbeel2007}. 

Most RL methods are based on learning the optimal control policy for the real system either directly, or indirectly. Indirect methods typically rely on learning a good approximation of the optimal action-value function underlying the MDP. The optimal policy is then indirectly obtained as the minimizer of the value-function approximation over the inputs $\vect a$. Direct RL methods, based on policy gradients, seek to adjust the parameters $\vect \theta$ of a given policy $\pi_{\vect\theta}$ such that it yields the best closed-loop performance when deployed on the real system. An attractive advantage of direct RL methods over indirect ones is that they are based on formal necessary conditions of optimality for the closed-loop performance of $\pi_{\vect\theta}$, and therefore asymptotically (for a large enough data set) guarantee the (possibly local) optimality of the parameters $\vect\theta$ \cite{Sutton1999, Silver2014}.

RL methods often rely on Deep Neural Networks (DNN) to carry the policy approximation $\pi_{\vect\theta}$. While effective in practice, control policies based on DNNs provide limited opportunities for formal verifications of the resulting closed-loop behavior, and for imposing hard constraints on the evolution of the state of the real system. The development of safe RL methods, which aims at tackling this issue, is currently an open field or research \cite{Garcia2015}. 

In this paper, we investigate the use of constrained parametric optimization problems to carry the policy approximation. The aim is to impose safety by means of hard constraints in the optimization problem. In that context, we investigate some straightforward options to build a safe stochastic policy, and discuss their shortcomings when using the stochastic policy gradient method. We then present an alternative approach, and propose tools to make its deployment computationally efficient, using the using primal-dual interior-point method and techniques from parametric Nonlinear Programming.

%Most RL methods require exploration, i.e., the inputs applied to the real system must differ from the policy $\vect\pi_{\vect\theta}$ in order to identify changes in the policy parameters ${\vect\theta}$  that can yield a higher closed-loop performance. Exploration is typically performed via stochastic disturbances of the policy. We will show in this paper that the presence of hard constraints distorts the statistics of the exploration, and that some corrections must in theory be introduced in the classic tools underlying the deterministic policy gradient method to account for this distortion. We propose computationally efficient tools to implement these corrections, based on parametric Nonlinear Programming techniques, and interior-point methods.

Robust Nonlinear Model Predictive Control (NMPC) is arguably an ideal candidate for forming the constrained optimization problem supporting the policy approximation. %Robust NMPC offers a pathway to treat systematically the safety requirements for dynamic systems, by enforcing the system safety explicitly.
%Nonlinear Model Predictive Control (NMPC) is a formal control method based on solving at every time instant an optimal control problem to generate the optimal control policy. The optimal control problem seeks to minimize a sum of stage costs $\ell(s,a)$ over a prediction horizon, subject to state trajectories provided by a model of the real system, and possibly state and input constraints to be respected. The optimal control problem then delivers an entire input and state sequence spanning the prediction horizon. Only the first input is applied to the real system, as the entire optimal control problem is solved again at the next time instant, based on the latest observations of the system evolution. 
Robust NMPC techniques provide safety guarantees on the closed-loop behavior of the system by explicitly accounting for the presence of (possibly stochastic) disturbances and model inaccuracies. A rich theoretical framework is available on the topic \cite{Mayne2014}. The policy parameters $\vect \theta$ then appear as parameters in the NMPC model(s), cost function and constraints. Updates in the policy parameters $\vect \theta$ are driven by the stochastic policy gradient method, increasing the NMPC closed-loop performance, and constrained by the requirement that the NMPC model inaccuracies are adequately accounted for in forming the robust NMPC scheme. For the sake of brevity and simplicity, we will detail these questions in the specific linear robust MPC case. The extension to the nonlinear case is arguably possible, but more complex.

This paper has a companion paper \cite{Gros2019b} treating the same problem in the context of the deterministic policy gradient approach. The two papers share some material and use similar techniques, but present very different theories. % allowing the deployment of the two policy gradient techniques is intrinsically different. 
Additionally~\cite{Zanon2019b} discusses the management of safety in RL using tube-based techniques.

The paper is structured as follows. Section \ref{sec:Background} provides some background material. Section \ref{sec:NMPCIntro} details the safe deterministic policy we use to build up the stochastic policy. Section \ref{sec:SafeStochasticPolicy} investigates several options to build safe stochastic policies from the deterministic policy approximation, discusses their shortcomings and proposes a computationally efficient alternative based on disturbed parametric NLPs. Section \ref{sec:Numerics} presents numerical tools for an efficient deployment of the stochastic policy gradient approach for the latter approach, using the primal-dual interior-point method and tools from parametric Nonlinear Programming. Section \ref{sec:SafeRLSteps} discusses a technique to ensure safety throughout the learning process, in the context of robust linear MPC. Section \ref{sec:Simulations} proposes an example of simulation using the principles developed in this paper.

%Section \ref{sec:ZeSection} establishes the fundamental result of the paper, showing that an (E)NMPC scheme based on the wrong model can, under some conditions, nonetheless deliver the optimal control policy $\pi_\star$. The Theorem establishing the result is fairly abstract, but it is illustrated in the LQR case and its application to (E)NMPC and robust (E)NMPC with scenario trees is detailed. Section \ref{sec:ENMPC} details the connection of the proposed theory to the fundamental concept of strict dissipativity underlying the stability theory of ENMPC, and details its consequences for using ENMPC as a parametrization for RL. Section \ref{eq:RLENMPC} details how RL methods can be deployed to tune (E)NMPC schemes. Section \ref{sec:Examples} proposes some illustrative examples.

%We ought to underline here that this paper shares strong connections with its companion paper \cite{GrosStochastic} treating the same problem in the context of the stochastic policy gradient approach.

\section{Background on Markov Decision Processes} \label{sec:Background}
In the following, we will consider that the dynamics of the real system are described as a Markov Chain, with state transitions having the underlying conditional probability density:
\begin{align}
\label{eq:State:Transition}
\mathbb{P}\left[\vect{s}_{+}\,|\,\vect{s},\vect{a}\right]
\end{align}
denoting the probability density of the state transition from the state-input pair $\vect{s}\in\mathbb{R}^n,\,\vect{a}\in\mathbb{R}^{n_{\vect a}}$ to a new state $\vect{s}_{+}\in \mathbb{R}^n$. We will furthermore consider (possibly) stochastic policies $\vect\pi$, taking the form of probability densities:
\begin{align}
\label{eq:StochPolicyDef}
{\pi}\left[\vect{a}\,|\,\vect{s}\right]
\end{align}
denoting the probability density of selecting a given input $\vect a$ when the system is in a given state $\vect s$. We should note here that a deterministic policy 
\begin{align}
\vect a = \vect\pi\left(\vect s\right)
\end{align}
can always be cast as a stochastic policy \eqref{eq:StochPolicyDef} by defining:
\begin{align}
\label{eq:Stoch:Det:Equivalence}
{\pi}\left[\vect{a}\,|\,\vect{s}\right] = \delta\left(\vect{a} - \vect{\pi}\left(\vect{s}\right)\right)
\end{align}
where $\delta$ is the Dirac function. Let us then consider the distribution of the Markov Chain resulting from the state transition \eqref{eq:State:Transition} and policy \eqref{eq:StochPolicyDef}:
\begin{align}
\mathbb{P}\left[\vect{s}_k\,|\,{\pi}\right] = \int \prod_{i=0}^{k-1} &\mathbb{P}\left[\vect{s}_{i+1}\,|\,\vect{s}_i,\vect{u}\right] \mathbb{P}\left[\vect{s}_0\right]{\pi}\left[\vect{a}_i\,|\,\vect{s}_i\right]\\ &\mathrm{d}\vect{s}_{0,\ldots,k-1}\mathrm{d}\vect{a}_{0,\ldots,k-1}\nonumber
\end{align}
where $\mathbb{P}\left[\vect{s}_0\right]$ denotes the probability distribution of the initial conditions $\vect s_0$ of the MDP. We can then define the discounted expected value under policy $\vect{\pi}$, which reads as:
\begin{align}
\mathbb{E}_{{{\pi}}}[\zeta\left(\vect{s},\vect{a}\right)] = \sum_{k=0}^\infty \int \gamma^k& \zeta(\vect{s}_k, \vect{a}_k)  \mathbb{P}\left[\vect{s}_k\,|\,{\pi}\right]{\pi}\left[\vect{a}_k\,|\,\vect{s}_k\right]\mathrm{d}\vect{s}_k\mathrm{d}\vect{a}_k
\end{align}
for any function $\zeta$. This definition can be easily extended for functions over state transitions, i.e. $\zeta\left(\vect{s}_+,\vect{s},\vect{a}\right)$.
%In the case the expected value is taken over a function depending on transitions, i.e. $\zeta\left(\vect{x}_+,\vect{x},\vect{u}\right)$, then the discounted expected value under policy $\vect{\pi}^\mathrm{s}$ is given by:
%\begin{align}
%&\mathbb{E}_{\tau_{\vect{\pi}^\mathrm{s}}}[\zeta\left(\vect{x}_+,\vect{x},\vect{u}\right)] = \sum_{k=0}^\infty \int \gamma^k \zeta(\vect{x}_{k+1},\vect{x}_k, \vect{u}_k)   \mathbb{P}\left[\vect{x}_k\,|\,\vect{\pi}^\mathrm{s}\right]\nonumber\\ &\qquad\qquad \cdot\mathbb{P}\left[\vect{x}_{k+1}\,|\,\vect{x}_k,\vect{u}_k\right] \vect{\pi}^\mathrm{s}\left[\vect{u}_k\,|\,\vect{x}_k\right]\mathrm{d}\vect{x}_k\mathrm{d}\vect{u}_k\mathrm{d}\vect{x}_{k+1}
%\end{align}
%In particular, the NMPC policy performance is assessed as the expected value of the stage cost $\ell$ under policy 
%\begin{align}
%\vect{\pi}\left[\vect{u}\,|\,\vect{x}\right] = \delta\left(\vect{u} - \vect{\pi}_{\vect{\theta}}\left(\vect{x}\right)\right)
%\end{align}
%i.e.:
%\begin{align}
%J\left(\vect{\pi}\right) = \mathbb{E}_{\tau_{\vect{\pi}}}[\ell\left(\vect{x},\vect{u}\right)] 
%\end{align}
In the following we will assume the local stability of the MDP under the selected policies. More specifically, we assume that $\pi$ is such that:% for a policy $\tilde{\vect \pi}^\mathrm{s}$ the following equality holds:
\begin{align}
\label{ass:Stability}
\lim_{\tilde{ \pi}\rightarrow{ \pi}} \mathbb{E}_{{\tilde{ \pi}}}\left[\zeta\right] = \mathbb{E}_{{{\pi}}}\left[\zeta\right], 
\end{align}
where the limit is taken in the sense of almost everywhere, and for any bounded function $\zeta$ such that both sides of the equality are finite. Assumption \eqref{ass:Stability} will allow us to draw equivalences between a policy and disturbances of that policy, which will be required in the RL context.

For a given stage cost function $L(\vect x,\vect u)$ and a discount factor $\gamma \in [0,1]$, the performance of policy $\pi$ is given by the discounted cost:
\begin{align}
\label{eq:Return}
J(\pi) = \mathbb{E}_{{ \pi}}\left[\, L \, \right]%\left.  L(\vect s_k,\vect a_k)\,\right|\, \vect a_k \sim  \pi\left[\, .\,|\,\vect s_k\right]\, \right]
\end{align}
The optimal policy associated to the MDP defined by the state transition \eqref{eq:State:Transition}, the stage cost $L$ and the discount factor $\gamma$ is then given by:
\begin{align}
\label{eq:OptimalPolicy}
\pi_\star =\mathrm{arg} \min_{ \pi}\, J(\pi)
\end{align}
It should be useful to underline here that, while \eqref{eq:OptimalPolicy} may have several (global) solutions, any fully observable MDP admits a deterministic policy $\vect\pi_\star$ among its solutions. The value function associated to a given policy $\pi$ is given by \cite{Bertsekas1995,Bertsekas1996,Bertsekas2007}:
%\begin{subequations}
\begin{align}
%Q_{\vect\pi}\left(\vect s,\vect a\right) &= L(\vect s,\vect a) + \gamma \mathbb{E}\left[V_{\vect\pi}(\vect s_{+})\,|\, \vect s,\, \vect a\right],  \label{eq:MDP:Qfunction:Generic}\\
V_{\vect\pi}\left(\vect s\right) &= \mathbb{E}_{\vect a\sim \pi[.|\vect x]}\left[ L(\vect s,\vect a) + \gamma \mathbb{E}\left[V_{\vect\pi}(\vect s_{+})\,|\, \vect s,\, \vect a\right]\right], \label{eq:Bellman:Policy:0}
\end{align}
%\end{subequations}
where the internal expected value in \eqref{eq:Bellman:Policy:0} is taken over state transitions \eqref{eq:State:Transition}. 
%The advantage function associated to a given policy is then given as:
%\begin{align}
%A_{\vect\pi}\left(\vect s,\vect a\right) = Q_{\vect\pi}\left(\vect s,\vect a\right) - V_{\vect\pi}\left(\vect s\right) 
%\end{align}
%and provides the value of using input $\vect u$ in a given state $\vect x$ compared to using the policy $\vect \pi$. In the case the policy $\vect \pi$ matches the optimal policy $\vect\pi_\star$ that minimizes \eqref{eq:Return}, then
%\begin{align}
%A_{\vect\pi}\left(\vect s,\vect a\right) = A_{\vect\pi_\star}\left(\vect s,\vect a\right) \geq 0,\quad \forall \vect x,\vect u
%\end{align}

%The optimal policy $\vect \pi^\star$ minimizing \eqref{eq:Return} is given by the Bellman optimality backup \cite{Berstekas}:
%\begin{subequations}
%\label{eq:Bellman:Generic:0}
%\begin{align}
%Q\left(\vect u,\vect x\right) &= \ell(\vect u,\vect x) + \gamma \mathbb{E}\left[V(\vect x_{+})\,|\, \vect x,\, \vect u\right],  \label{eq:MDP:Qfunction:Generic}\\
%V\left(\vect x\right) &= Q\left(\vect \pi(\vect x),\,\vect x\right), \\
%\vect \pi^\star(\vect x) &=  \mathrm{arg}\min_{\vect u} \,\, Q\left(\vect u,\,\vect x\right), \label{eq:MDP:Policy:Generic}
%\end{align}
%\end{subequations}
%For most problems of practical interest, solving the Bellman optimality backup \eqref{eq:Bellman:Generic:0} is overly expensive.

\subsection{Stochastic policy gradient} \label{sec:StochPolicyGradient}

In most cases, the optimal policy $\vect\pi_\star$ cannot be computed. It is then useful to consider a stochastic approximations $\pi_{\vect\theta}$ of the optimal policy, carried by a (possibly large) set of parameters $\vect\theta$. The optimal parameters $\vect\theta_\star$ are then given by:
\begin{align}
\vect \theta_\star = \mathrm{arg}\min_{\vect\theta}\, J(\pi_{\vect\theta}) \label{eq:OptParameters}
\end{align}
The policy gradient $\nabla_{{\theta}}\, J({\pi}_{\vect{\theta}}) $ associated to the stochastic policy $\pi_{\vect{\theta}}$ can be obtained using various actor-critic methods, such as e.g. \cite{Sutton1998,Sutton1999}: 
\begin{align}
\nabla_{{\theta}}\, J({\pi}_{\vect{\theta}}) %&= \mathbb{E}_{\tau_{\vect{\pi}_{\vect{\theta}}}}\left[\nabla_{{\theta}}\log \vect{\pi}_{\vect{\theta}}\, A_{\vect{\pi}_{\vect{\theta}}}\right], 
\label{eq:StochasticPiGradient}
&= \mathbb{E}_{{\vect{\pi}_{\vect{\theta}}}}\left[\nabla_{{\theta}}\log {\pi}_{\vect{\theta}}\, \delta^V_{{\pi}_{\vect{\theta}}}\right], %\nonumber
\end{align}
where 
\begin{align}
\label{eq:deltaV}
\delta^V_{\pi_{\vect{\theta}}}= L \left(\vect s,\vect a\right) + \gamma V_{{\pi}_{\vect{\theta}}}\left(\vect s_+\right)  - V_{{\pi}_{\vect{\theta}}}\left(\vect s\right)
\end{align}
%or any variant of it (e.g. based on the value function, action-value function, etc.).
The value function $V_{{\pi}_{\vect{\theta}}}$ %and advantage function $A_{\vect{\pi}_{\vect{\theta}}}$ 
in \eqref{eq:deltaV} is formally given by \eqref{eq:Bellman:Policy:0}, but 
%built from the action-value function $Q_{\vect{\pi}_{\vect{\theta}}}$, which is 
typically approximated via a parametrized value function approximation and computed via Temporal-Difference (TD) techniques or Monte-Carlo techniques \cite{Sutton1998}. Note that is is fairly common in RL to generate the stochastic policy $\pi_{\vect\theta}$ as a disturbed version of a deterministic policy $\vect\pi_{\vect\theta}$ by, e.g., adding a simple stochastic disturbance to $\vect\pi_{\vect\theta}$.

%via a compatible function approximation $\hat A^{\vect{w}}_{\vect{\pi}_{\vect{\theta}}}$ of the form \cite{Silver}:
%\begin{align}
%\hat A^{\vect{w}}_{\vect{\pi}_{\vect{\theta}}} = \vect{w}^\top \nabla_{{\theta}}\log \vect{\pi}_{\vect{\theta}}.
%\end{align}
%where $\vect w \in\mathbb{R}^{n_{\vect\theta}}$ are parameters to be estimated, typically via Temporal-Difference (TD) techniques \cite{Suton}:
%\begin{align}
%\label{eq:A:TDEstimation}
%\vect{w} =\mathrm{arg}\min_{\vect{w}}\, \frac{1}{2}\mathbb{E}_{\tau_{\vect{\pi}_{\vect{\theta}}}}\left[\left(Q_{\vect{\pi}_{\vect{\theta}}} - \hat V_{{\pi}_{\vect{\theta}}} - \hat A^{\vect{w}}_{\vect{\pi}_{\vect{\theta}}}   \right)^2\right],
%\end{align}
%supported by the baseline value function estimation 
%\begin{align}
%\hat V_{{\pi}_{\vect{\theta}}}\approx V_{{\pi}_{\vect{\theta}}},
%\end{align}
%also computed via TD techniques.

%In order for the baseline estimation error $\hat V_{{\pi}_{\vect{\theta}}}\neq V_{{\pi}_{\vect{\theta}}} $ to not affect the estimation of $\vect{w}$, the following condition must hold:
%\begin{align}
%\label{eq:UnbiasedConditionStochastic}
%\mathbb{E}_{\tau_{\vect{\pi}_{\vect{\theta}}}}\left[\nabla_{{\theta}}\log \vect{\pi}_{\vect{\theta}}\left(V_{{\pi}_{\vect{\theta}}}-\hat V_{{\pi}_{\vect{\theta}}}\right)\right] = 0.
%\end{align}
%However, since the baseline error is purely a function of the state, condition \eqref{eq:UnbiasedConditionStochastic} holds by construction \cite{Suton}. 

In order to deploy the stochastic policy, ${\pi}_{\vect{\theta}}$ needs to be sampled, to produce realizations of the inputs $\vect a$ to be deployed on the system. Moreover, in order to compute the policy gradient, evaluations of the gradient of the policy score function:
\begin{align}
\nabla_{{\theta}}\log {\pi}_{\vect{\theta}} ={\pi}_{\vect{\theta}}^{-1}\nabla_{\vect{\theta}}{\pi}_{\vect{\theta}}
\end{align}
are required in computing \eqref{eq:StochasticPiGradient}.
%and evaluate the policy gradient \eqref{eq:StochasticPiGradient} efficiently, one needs to build a stochastic policy $\vect{\pi}_{\vect{\theta}}$ that is inexpensive to both sample from and evaluate. Indeed, sampling is required in order to generate realizations of the inputs $\vect a$ to be deployed on the system, and evaluations of the policy are required in computing \eqref{eq:A:TDEstimation} and \eqref{eq:StochasticPiGradient}. 

%In order to illustrate this statement, the next subsection details a very natural approach to generate $\vect{\pi}_{\vect{\theta}}$, such that $\vect{\pi}_{\vect{\theta}}$ is fairly inexpensive to sample from, but unfortunately very expensive to evaluate.

%
%A difficulty when deploying a stochastic policy gradient approach under safety constraints on continuous action spaces is that the stochastic policy 
%
% is the problem of forming a stochastic policy that generates safe input samples while being fairly cheap to evaluate, even for non-trivial safety constraints. Indeed, generating safe input samples requires working with a distribution whose support satisfies the safety constraints. While it is fairly straightforward to draw safe samples, evaluating the stochastic policy on a given draw can be highly difficult for continuous input spaces (or high-dimensional discrete input spaces), and/or a non-trivial safety constraints. 
 
% This statement is illustrated in the following subsection. Section \ref{sec:IPsampling} proposes a solution to that problem using an Interior-Point approach.

\subsection{Safe set} \label{sec:SafeSet}
In the following, we will assume the existence of a (possibly) state-dependent \textit{safe set} labelled $\mathbb{S}\left(\vect s\right)\subseteq \mathbb{R}^{n_{\vect a}}$, subset of the input space. The notion of safe set will be used here in the sense that any input selected such that $\vect a\in \mathbb{S}\left(\vect s\right)$ yields safe future trajectories with a unitary probability. The construction of the safe set is not the object of this paper. However, we can nonetheless propose some pointer to how such a set is constructed in practice.

Let us consider the constraints $\vect h_\mathrm{s}\left(\vect s,\vect a\right)\leq 0$ describing at any given future time $i$ the subset of the state-input space deemed feasible and safe. Constraints $\vect h$ can include pure state constraints, describing the safe states, pure input constraints, describing typically actuators limitations, and mixed constraints, where the states and inputs are mixed. For the sake of simplicity, we will assume in the following that $\vect h_\mathrm{s}$ is convex.

%A practical definition of the safe set can then be constructed via reachability analysis \cite{}. Consider the set $\vect{X}_+\left(\vect{s},\vect{a}\right)$, support of the state transition \eqref{eq:State:Transition}, i.e.
%\begin{align}
%\label{eq:StateTransition:Support}
%	\mathbb{P}\left[\,\vect{s}_+\in \vect{X}_+\left(\vect{s},\vect{a}\right)\,\,|\,\, \vect{s},\vect{a}\,\right] = 1.
%\end{align}
%Let us then consider the recursion
%\begin{align}
%\mathbb{S}_{k-1} = \left\{\, \vect s\,\,|\,\,\exists\, \vect a\quad \mathrm{s.t.}\quad \vect h\left(\vect s,\vect a\right)\leq 0,\,\,\,\,  \vect X_+\left(\vect s,\vect a\right)\subseteq\mathbb{S}_{k} \right\}
%\end{align}
%with the boundary condition:
%\begin{align}
%\mathbb{S}_\infty = \left\{\, \vect s\,\,|\,\,\exists\, \vect a\quad \mathrm{s.t.}\quad  \vect h\left(\vect s,\vect a\right)\leq 0\right\}
%\end{align}
%The safe set can then be defined from the limit of this recursion as:
%\begin{align}
%\label{eq:SafeSet:DefintionViaReachability}
%\mathbb{S}\left(\vect s\right) = \left\{\,\vect a\,\,|\,\, \vect h\left(\vect s,\vect a\right)\leq 0,\,\,\,\,  \vect X_+\left(\vect s,\vect a\right)\subseteq\mathbb{S}_{0} \right\}
%\end{align}  
%If the support $\vect{X}_+\left(\vect{s},\vect{a}\right)$ of the probability density \eqref{eq:State:Transition} is known, or if an outer approximation is available, the construction of the safe set $\mathbb{S}\left(\vect s\right)$ using \eqref{eq:StateTransition:Support}-\eqref{eq:SafeSet:DefintionViaReachability} is viable for simple problems, but is often intractable for more complex ones.

A common approach to build practical or inner approximations of the safe set $\mathbb{S}\left(\vect s\right)$ is via verifying the safety of an input $\vect a$ explicitly over a finite horizon via predictive control techniques. This verification is based on forming the support of the Markov Process distribution over time, starting from a given state-input pair $\vect s,\vect a$. Consider the set $\vect{X}_+\left(\vect{s},\vect{a}\right)$, support of the state transition \eqref{eq:State:Transition},
\begin{align}
\label{eq:StateTransition:Support}
\vect{X}_+\left(\vect{s},\vect{a}\right) = \left\{\,\left.\vect s_+\,\,\right |\,\, \mathbb{P}\left[\,\vect s_+\,|\vect s,\vect a\right] > 0\,\right\}
%\mathbb{P}\left[\,\vect{s}_+\in \vect{X}_+\left(\vect{s},\vect{a}\right)\,\,|\,\, \vect{s},\vect{a}\,\right] = 1,%\quad 	\mathbb{P}\left[\,\vect{s}_+\notin \vect{X}_+\left(\vect{s},\vect{a}\right)\,\,|\,\, \vect{s},\vect{a}\,\right] = 0.
\end{align}
Labelling $\vect{X}_k(\vect{s},\vect a,\vect{\pi}^{\mathrm{s}})$ the support of the state of the Markov Process at time $k$, starting from $\vect s,\vect a$ and evolving under policy $\vect{\pi}^{\mathrm{s}}$, the set $\vect{X}_k$ is then given by the recursion:
\begin{align}
\label{eq:SetDispersion}
\vect{X}_k(\vect{s},\vect a,\vect{\pi}^{\mathrm{s}}) &= \vect{X}_+\left(\vect{X}_{k-1},\vect{\pi}^{\mathrm{s}}(\vect{X}_{k-1})\right),
\end{align}
with the boundary condition $ \vect{X}_1 =\vect{X}_+\left(\vect{s},\vect{a}\right)$. An input $\vect a$ is in the safe set $\mathbb{S}\left(\vect s\right)$ if $\vect h_\mathrm{s}\left(\vect s,\vect a\right)\leq 0$ and if there exist a deterministic policy $\vect\pi^\mathrm{s}$ such that
\begin{align}
\vect h_\mathrm{s}\left(\vect s_k,\vect\pi^s\left(\vect s_k\right)\right)\leq 0, \quad \forall\, \vect s_k\in \vect{X}_k(\vect{s},\vect a,\vect{\pi}^{\mathrm{s}}), 
\end{align}
for all $k\geq 1$. This verification is typically performed in practice via tube-based approaches, polynomial chaos, or direct approximations of the set $\vect{X}_k$ via e.g. ellipsoids or polytopes. In that context, policy $\vect{\pi}^{\mathrm{s}}$ is typically selected a priori to stabilize the system dynamics, and possibly optimized to minimize the size of the sets $\vect{X}_k$. %In this paper, we will adopt a scenario tree approach, whereby 
\subsection{Safe stochastic policy}
In this paper, we will consider safe, stochastic policies $\pi_{\vect\theta}$, which we will label 
\begin{align}
{\pi}_{\vect{\theta}}\left[\vect a\,|\, \vect s\right]
\end{align}
We will build ${\pi}_{\vect{\theta}}$ from a deterministic policy $\vect\pi_{\vect\theta}$ based on a constrained optimization scheme such that the support of ${\pi}_{\vect{\theta}}$ is limited to the safe set $\mathbb{S}(\vect s)$, i.e. such that 
\begin{align}
\mathbb{P}\left[\vect a\notin \mathbb{S}(\vect s)\,|\,\vect a \sim {\pi}_{\vect{\theta}}\left[\vect a\,|\, \vect s\right] \,\right] = 0
\end{align}
Unfortunately, when the support of the stochastic policy ${\pi}_{\vect{\theta}}$ must be restricted within the given, possibly non-trivial safe set $\mathbb{S}(\vect s)$, it is not always straightforward to build a stochastic policy ${\pi}_{\vect{\theta}}$ that is at the same time inexpensive to sample from and to evaluate. 

 There are clearly several approaches to generate random inputs that are in the safe set $\mathbb{S}(\vect s)$ with unitary probability, but we will focus here on techniques that require a limited amount of computations, so as to make them real-time feasible.

\section{Optimization-based safe policy} \label{sec:NMPCIntro}

In this paper, we will consider parametrized deterministic policies $\vect{\pi}_\theta\approx \vect\pi_\star$ based on parametric optimization problems subject to safe stochastic disturbances. Before detailing the stochastic aspect, let us detail first the constrained optimization problems. We will consider parametrized deterministic policies $\vect{\pi}_\theta$ based on parametric Nonlinear Programs (NLPs), and more specifically based on robust NMPC schemes. This approach is formally justified in \cite{Gros2018}. More specifically, we will consider a policy approximation 
\begin{align}
\vect\pi_{\theta} = \vect{u}^\star_0\left(\vect{s},\vect{\theta}\right), \label{eq:PolicyFromNLP}
\end{align}
 where $\vect{u}^\star_0\left(\vect{s},\vect{\theta}\right)$ is the first $n_{\vect a}$ entries of $\vect{u}^\star\left(\vect{s},\vect{\theta}\right)$ generated by the parametric NLP:
\begin{subequations}
\label{eq:Generic:NLP}
\begin{align}
\vect{u}^\star\left(\vect{s},\vect{\theta}\right) = \mathrm{arg}\min_{\vect{u}}&\quad \Phi(\vect{x},\vect{u},\vect{\theta}) \\
\mathrm{s.t.}&\quad \vect f\left(\vect x,\vect u,\vect s,\vect{\theta}\right) = 0, \label{eq:DynamicConstraints}\\
&\quad \vect{h}\left(\vect{x},\vect{u},\vect{\theta}\right) \leq 0. \label{eq:SafetyConstraints}
\end{align}
\end{subequations}
We will then consider that the safety requirement $\vect \pi_{\vect\theta}(\vect s) \in\mathbb{S}(\vect s)$ is imposed via the constraints \eqref{eq:DynamicConstraints}-\eqref{eq:SafetyConstraints}. A special case of \eqref{eq:Generic:NLP} is an optimization scheme in the form:
\begin{subequations}
\label{eq:RobustNMPC:Policy:Constraints:Static}
\begin{align}
\vect u_0^\star\left(\vect s,\vect{\theta}\right) = \mathrm{arg}\min_{\vect u_0}&\quad \Phi(\vect s,\vect u_0,\vect{\theta}) \\
\mathrm{s.t.}&\quad \vect{h}\left(\vect s,\vect u_0,\vect{\theta}\right) \leq 0, \label{eq:SafetyConstraints:Static}
\end{align}
\end{subequations}
where $\vect{h}\leq0$ ought to ensure that $\vect \pi_{\vect\theta}(\vect s) = \vect u_0^\star\left(\vect s,\vect{\theta}\right) \in\mathbb{S}\left(\vect s\right)$. 

While most of the discussions in this paper will take place around the general formulation \eqref{eq:Generic:NLP}, a natural approach to formulate constraints \eqref{eq:DynamicConstraints}-\eqref{eq:SafetyConstraints} such that policy \eqref{eq:PolicyFromNLP} is safe is to build \eqref{eq:Generic:NLP} using robust (N)MPC techniques. %We detail that next.

 \subsection{Policy approximation based on robust NMPC}
The imposition of safety constraints can be treated via robust NMPC approaches. Robust NMPC can take different forms \cite{Mayne2014}, all of which can be eventually cast in the form \eqref{eq:Generic:NLP}. One form of robust robust NMPC schemes is based on scenario trees~\cite{Scokaert1998}, which take the form: 
\begin{subequations}
\label{eq:RobustNMPC:Policy:Constraints:MPC}
\begin{align}
\vect{u}^\star\left(\vect s,\vect{\theta}\right) &=\nonumber\\ \mathrm{arg}\min_{\vect{u}}&\,\, \sum_{j=1}^{N_{\mathrm M}} \left(V_j(\vect x_{j,N},\vect\theta) +  \sum_{k=0}^{N-1} \ell_j(\vect x_{j,k},\vect u_{j,k},\vect\theta)\right)\\
\mathrm{s.t.}&\,\, \vect x_{j,k+1} = \vect F_j\left(\vect x_{j,k},\vect u_{j,k},\vect \theta\right),\,\,\, \vect x_{j,0} = \vect s, \label{eq:Dynamics:MPC:Robust}\\
&\,\,  \vect{h}^\mathrm{s}\left(\vect{x}_{j,k},\vect{u}_{j,k},\vect \theta\right) \leq 0, \label{eq:SafetyConstraints:MPC:Robust} \\
&\,\,  \vect{e}\left(\vect{x}_{j, N},\vect \theta\right) \leq 0, \label{eq:TerminalConstraints:MPC:Robust} \\
%&\,\, \vect u_{j,0} = \vect u_{i,0},\quad \forall\, i,j = 0,\ldots N_{\mathrm M},\\
&\,\,\, \vect N\left(\vect u\right )= 0, \label{eq:NonAnticipativity}%,\quad \forall\, i,j = 0,\ldots N_{\mathrm M},\\
%&\,\,\, \vect u_{j,k} = \vect\pi^\mathrm{s}\left(\vect x_{j,k},\vect u_{0,k},\vect x_{0,k}\right),\,\, \forall\, k=1,\ldots N,
%&\,\, \vect u_{j,k} = \vect\pi^\mathrm{s}\left(\vect x_{j,k}, \vect x_{0,k}, \vect u_{0,k} \right),\, \forall\, k\geq 1,\, j\geq 1,
\end{align}
\end{subequations}
%\seb{I would still like to refine this... I'd be happy to get some help / discuss}
where $\vect F_{1,\ldots,N_{\mathrm M}}$ are the $N_{\mathrm M}$ different models used to support the uncertainty, while $\vect F_0$ is a nominal model supporting the NMPC scheme. Trajectories $\vect x_{j,k}$ and $\vect u_{j,k}$ for $j=1,\ldots,N_\mathrm{M}$ are the different models trajectories and the associated inputs. Functions $\ell_{1,\ldots,N_{\mathrm M}}$, $V_{1,\ldots,N_{\mathrm M}}$ the (possibly different) stage costs and terminal costs applying to the different models. The \textit{non-anticipativity constraints} \eqref{eq:NonAnticipativity} support the scenario-tree structure.
%Model $j=0$ plays a special role in \eqref{eq:RobustNMPC:Policy:Constraints:MPC}, and is labelled nominal model, see Section \ref{sec:SafeTree}.
For a given state $\vect s$ and parameters $\vect \theta$, the NMPC scheme \eqref{eq:RobustNMPC:Policy:Constraints:MPC} delivers the input profiles 
\begin{align}
\vect{u}_j^\star\left(\vect s,\vect{\theta}\right) = \left\{\vect u_{j,0}^\star\left(\vect s,\vect{\theta}\right),\ldots, \vect{u}_{j,N}^\star\left(\vect s,\vect{\theta}\right)\right\},\end{align}
with $\vect{u}_{j,i}^\star\left(\vect s,\vect{\theta}\right)\in\mathbb{R}^{n_{\vect a}}$, and \eqref{eq:NonAnticipativity} imposes
\begin{align}
\vect u_0^\star\left(\vect s,\vect{\theta}\right) := \vect u_{i,0}^\star\left(\vect s,\vect{\theta}\right) = \vect u_{j,0}^\star\left(\vect s,\vect{\theta}\right),\quad\forall i,j.
\end{align}
As a result, the NMPC scheme \eqref{eq:RobustNMPC:Policy:Constraints:MPC} generates a parametrized deterministic policy according to:
\begin{align}
\label{eq:Policy0}
\vect{\pi}_\theta\left(\vect s\right) = \vect u_0^\star\left(\vect{s},\vect{\theta}\right) \,\in\,\mathbb{R}^{n_{\vect a}}.
\end{align}
Policy $\vect\pi^\mathrm{s}$ is implicitly deployed in \eqref{eq:RobustNMPC:Policy:Constraints:MPC} via the scenario tree. {If the dispersion set $\vect X_+$ is known, the multiple models $\vect F_{1,\ldots , N_\mathrm{M}}$ and terminal constraints \eqref{eq:TerminalConstraints:MPC:Robust} can be chosen such that the robust NMPC scheme \eqref{eq:RobustNMPC:Policy:Constraints:MPC} delivers $\vect{\pi}_\theta\left(\vect s\right)\in\mathbb{S}\left(\vect s\right)$. Unfortunately, this selection can be difficult in general. We turn next to the robust linear MPC case, where this construction is much simpler.}

\subsection{Safe robust linear MPC} \label{sec:SafeTree}
Exhaustively discussing the construction of the safe scenario tree in \eqref{eq:RobustNMPC:Policy:Constraints:MPC} for a given dispersion set $\vect X_+\left(\vect s,\vect a\right)$ is beyond the scope of this paper. The process can be fairly involved, and we refer to \cite{Scokaert1998,Bernardini2009} for detailed discussions. For the sake of brevity, we will focus on the linear MPC case, whereby the MPC models $\vect F_{1,\ldots,N_{\mathrm M}}$ and policy $\vect\pi^\mathrm{s}$ are linear. 

Let us consider the following outer approximation of  the dispersion set $\vect X_+$:
\begin{align}
\label{eq:DispersionApprox}
\vect X_+\left(\vect s,\vect a\right) \subseteq \vect F_0\left(\vect s,\vect a,\vect \theta\right) + \vect W,\quad \forall \, \vect s,\vect a
\end{align}
where we use a linear nominal model $\vect F_0$ and a polytope $\vect W$ of vertices $\vect W^{1,\ldots,N_{\mathrm M}}$ that can be construed as the extrema of a finite-support process noise, and which can be part (or functions of) the MPC parameters $\vect\theta$. For the sake of simplicity, we assume that $\vect W$ is independent of the state-input pair $\vect s,\vect a$. The models $\vect F_{1,\ldots,N_\mathrm M}$ can then be built based using:
\begin{align}
\vect F_i = \vect F_0 + \vect W^i,\quad i=1\ldots N_{\mathrm M}
\end{align}
and using the linear policy:
\begin{align}
\vect\pi^\mathrm{s}\left(\vect x_{j,k},\vect u_{0,k},\vect x_{0,k}\right) = \vect u_{0,k} - K\left(\vect x_{j,k}-\vect x_{0,k}\right)
\end{align}
where matrix $K$ can be part (or function of) the MPC parameters $\vect\theta$. One can then verify by simple induction that:
\begin{align}
\vect{X}_k(\vect{s},\vect a,\vect{\pi}^{\mathrm{s}}) \subseteq \mathrm{Conv}\left(\vect x_{1,k},\ldots, \vect x_{N_{\mathrm M},k}\right),
\end{align}
for $k=0,\ldots,N+1$, where $ \mathrm{Conv}$ is the convex hull of the set of points $\vect x_{1,k},\ldots, \vect x_{N_{\mathrm M},k}$ solution of the MPC scheme \eqref{eq:RobustNMPC:Policy:Constraints:MPC}. The terminal constraints \eqref{eq:TerminalConstraints:MPC:Robust} ought then be constructed as, e.g., via the Robust Positive Invariant set corresponding to $\vect\pi^\mathrm{s}$ in order to establish safety beyond the MPC horizon. For $\vect{h}^\mathrm{s}$ convex, the MPC scheme \eqref{eq:RobustNMPC:Policy:Constraints:MPC} delivers safe inputs \cite{Mayne2014,Kolmanovsky1998}.
%\textcolor{blue}{\textit{Explain more...}} \mario{a bit too quick above here} \seb{we should discuss what aspect needs to be extended.}

When the dispersion set $\vect X_+\left(\vect s,\vect a\right) $ can only be inferred from data, condition \eqref{eq:DispersionApprox} arguably translates to \cite{Bertsekas1971a}:
\begin{align}
\label{eq:DispersionApprox:Data}
\vect s_{k+1} - \vect F_0\left(\vect s_k,\vect a_k,\vect \theta\right) \in  \vect W,\quad \forall \left(\vect s_{k+1} ,\vect s_k,\vect a_k\right)\in\mathcal D,%k = 0,\ldots,N_\mathrm{D}
\end{align}
where $\mathcal D$ is the set of $N_{\mathcal{D}}$ observed state transitions. Condition \eqref{eq:DispersionApprox:Data} translates into a sample-based condition on the admissible parameters $\vect \theta$, i.e., it speficies the parameters that are safe \textit{with respect to the state transitions observed so far}. Condition \eqref{eq:DispersionApprox:Data} tests whether the points $\vect s_{k+1} - \vect F_0\left(\vect s_k,\vect a_k,\vect \theta\right)$ are in the polytope $\vect W$, which can be easily translated into a set of algebraic constraints imposed on $\vect\theta$. This observation will be used in Section \ref{sec:SafeTree} to build a safe RL-based learning.

We ought to underline here that building $\vect F_0,\,\vect W$ based on \eqref{eq:DispersionApprox:Data} ensures the safety of the robust MPC scheme \eqref{eq:RobustNMPC:Policy:Constraints:MPC} only for an infinitely large, and sufficiently informative data set $\mathcal D$. In practice, using a finite data set entails that safety is ensured with a probability less than 1. The quantification of the probability of having a safe policy for a given, finite data set $\mathcal D$ is beyond the scope of this paper, and is arguably best treated by means of the Information Field Theory \cite{Ensslin2013}. The extension of the construction of a safe MPC presented in this section to the general NMPC case is theoretically feasible, but can be computationally intensive in practice. This aspect of the problem is beyond the scope if this paper.

\section{Safe stochastic policies} \label{sec:SafeStochasticPolicy}
In this section, we discuss first two intuitively appealing methods to generate safe stochastic policies from $\vect\pi_{\vect\theta}$, see Sections \ref{sec:Resampling} and \ref{sec:IPsoftmax}, and detail their computational shortcomings in the context of NMPC-based RL discussed in this paper. We present then an alternative approach in Section \ref{sec:IPsampling}.

\subsection{Safe stochastic policy via resampling} \label{sec:Resampling}
%\textcolor{red}{This could be sacrified for space}

Let us discuss first a very natural approach to generating a safe stochastic policy, based on re-generating a random input $\vect a$ until it is in the safe set $\mathbb{S}(\vect s)$. This can, e.g., be achieved by the trivial re-sampling Algorithm~\ref{alg:resampling}, where $\varrho\left( .\,|\, \vect{\pi}_\theta\left(\vect s\right)\right)$ is a probability density centred at $\vect{\pi}_\theta\left(\vect s\right)$.
\begin{algorithm}[t]
\caption{Resampling}
\label{alg:resampling}
 \DontPrintSemicolon
  \SetKwInOut{Input}{Input}
\Input{State $\vect s$, conditional density $\varrho(.\,|\,.)$}
Set \textit{sample} = true\\
\While{sample}{
Draw $\vect a\sim\varrho\left(.\,|\,\vect{\pi}^\mathrm{d}_\theta\left(\vect s\right)\right)$ \\
\If{$\vect a\in \mathbb{S}(\vect s)$}{
\textit{sample} = false
}
}
\Return $\vect a$
\end{algorithm}
One can, e.g., choose for $\varrho(.\,|\,.)$ a Normal distribution centered at $\vect{\pi}_\theta\left(\vect s\right)$, i.e., 
\begin{align}
\vect a\sim \mathcal{N}\left(\vect{\pi}_\theta\left(\vect s\right),\,\Sigma\right),
\end{align}
Verifying the condition $\vect a\in \mathbb{S}(\vect s)$ can then be done via classic optimization techniques, where one verifies the feasibility of the constraints \eqref{eq:SafetyConstraints} when selecting $\vect u_0 =\vect a$ in \eqref{eq:Generic:NLP} according to the proposed $\vect a$.

One can verify that the resulting stochastic policy ${\pi}_{\vect{\theta}}\left[\vect a\,|\,\vect s\right]$ takes the probability density:
\begin{align}
{\pi}_{\vect{\theta}}\left[\vect a\,|\,\vect s\right] = \frac{\varrho\left(\vect a\,|\,\vect{\pi}_\theta\left(\vect s\right)\right)}{\mu_{\varrho}\left(\mathbb{S}(\vect s)\right)},
\end{align}
where $\mu_{\varrho}\left(\mathbb{S}(\vect s)\right)$ is the measure of density $\varrho$ over $\mathbb{S}(\vect s)$, i.e.,
\begin{align}
\label{eq:Measure}
\mu_{\varrho}\left(\mathbb{S}(\vect s)\right) = \int_{\mathbb{S}(\vect s)} \varrho\left(\vect a\,|\,\vect{\pi}_\theta\left(\vect s\right)\right) \mathrm{d}\vect a.
\end{align}
%An advantage of the resampling approach is then that one can directly select the safe stochastic policy ${\pi}_{\vect{\theta}}\left[\vect a\,|\,\vect s\right]$ over $\mathbb{A}(\theta)$, up to the normalization factor $\mu_{\varrho}\left(\mathbb{A}(\theta),X_0\right)$.
%Unfortunately \eqref{eq:Measure} can be very expensive to evaluate accurately, as it typically requires sampling techniques. 
We observe then that the gradient of the score function required in calculating \eqref{eq:StochasticPiGradient} reads as:
\begin{align}
\nabla_{{\theta}}\log {\pi}_{\vect{\theta}}\left[\vect a\,|\,\vect s\right] =& \nabla_{\vect{\theta}}\log \varrho\left(\vect a\,|\,\vect{\pi}_\theta\left(\vect s\right)\right) \nonumber \\
&\hspace{4em}-\nabla_{\vect{\theta}}\log \mu_{\varrho}\left(\mathbb{S}(\vect s)\right),\label{eq:Resampling:GradScore}
\end{align}
such that
\begin{align}
\label{eq:ScoreOfMeasure}
\nabla_{\vect{\theta}}\log \mu_{\varrho}\left(\mathbb{S}(\vect s)\right) = -\frac{\nabla_{\vect{\theta}} \mu_{\varrho}\left(\mathbb{S}(\vect s)\right) }{\mu_{\varrho}\left(\mathbb{S}(\vect s)\right)}.
\end{align}
%While evaluating $\mu_{\varrho}\left(\mathbb{S}(\vect s)\right)$ accurately can be arguably done at limited computational expenses via sampling techniques, if $\varrho$ is inexpensive to sample from.

A difficulty arising in the re-sampling approach is that if the safe set $\mathbb{S}(\vect s)$ is not trivial, evaluating \eqref{eq:ScoreOfMeasure} can only be done via sampling techniques. Sampling techniques are computationally efficient only if verifying the condition $\vect a\in \mathbb{S}(\vect s)$ is inexpensive. This is unfortunately not the case in the NMPC context, where verifying $\vect a\in \mathbb{S}(\vect s)$ requires solving an NLP, or at least a feasibility problem.
%as the safety set $\mathbb{S}(\vect s)$ is defined by \eqref{eq:SDefinition}, verifying the latter conditions is unfortunately expensive. 
Furthermore, evaluating the gradient of the measure $\nabla_{\vect{\theta}} \mu_{\varrho}\left(\mathbb{S}(\vect s)\right)$ via sampling is in general even more difficult. 
\subsection{Safe stochastic policy via softmax} \label{sec:IPsoftmax}
%\textcolor{red}{This could be sacrified for space}
We consider next a classic approach in RL to generate stochastic policies, based on the softmax approach, but adapted to the optimization-based policy approximation. Consider the following modification of \eqref{eq:Generic:NLP}, based on the primal interior-point method \cite{Biegler2010}, using a logarithmic barrier:
\begin{subequations}
\label{eq:RobustNMPC:Policy:Softmax}
\begin{align}
\Phi^\star_\tau\left(\vect{a},\vect s,\vect{\theta}\right) =&\nonumber\\\min_{\vect{u}}&\,\,\, \Phi(\vect x,\vect{u},\vect{\theta}) - \tau \sum_i\log\left(\vect{h}_i\left(\vect x,\vect{u},\vect{\theta}\right)\right)\\
\mathrm{s.t.}&\,\,\, \vect f\left(\vect x,\vect u,\vect s,\vect{\theta}\right) = 0, \label{eq:eq:RobustNMPC:Policy:Softmax:Dynamics} \\
&\,\,\, \vect u_0 = \vect a. \label{eq:eq:RobustNMPC:Policy:Softmax:InputEmbedding}
\end{align}
\end{subequations}
Infeasible inputs $\vect a$ ought to be treated by assigning an infinite value to $\Phi^\star_\tau$. A stochastic policy can then be defined as a softmax \cite{Sutton1998} over the cost of \eqref{eq:RobustNMPC:Policy:Softmax} , i.e.:
\begin{align}
\pi_{\vect\theta}\left[ \, \vect a\,|\,\vect s\,\right] \propto e^{-\Phi_\tau^\star\left(\vect{a},\vect s,\vect{\theta}\right) }. \label{eq:MPC:Softmax}
\end{align}
We then observe that the gradient of the score function can be obtained via NLP sensitivity techniques \cite{Nocedal2006} and reads as
\begin{align}
\nabla_{\vect\theta} \log  \pi_{\vect\theta}\left[ \, \vect a\,|\,\vect s\,\right] &= - \nabla_{\vect\theta} \Phi_\tau^\star\left(\vect{a},\vect s,\vect{\theta}\right) = - \nabla_{\vect\theta} \mathcal L,
\end{align}
where $\mathcal L$ is the Lagrange function associated to \eqref{eq:RobustNMPC:Policy:Softmax}, i.e.,
\begin{align}
\hspace{-2pt}\mathcal L(\vect x,\vect{u},\vect\lambda,\vect\mu,\vect{\theta}) &= \Phi - \tau \sum_i\vect{h}_i+ \vect \lambda_0^\top \left(\vect u_0 - \vect a\right) +\vect\lambda^\top\vect f,
\end{align}
and $\vect \lambda,\,\vect \lambda_0$ are the multipliers associated to constraint \eqref{eq:eq:RobustNMPC:Policy:Softmax:Dynamics} and \eqref{eq:eq:RobustNMPC:Policy:Softmax:InputEmbedding}, respectively. 

Sampling the softmax policy \eqref{eq:MPC:Softmax} requires, in general, Importance Sampling techniques like the Metropolis-Hastings Algorithm  (MHA), allowing one to sample an arbitrary continuous distribution. The difficulty with such techniques is that they typically require a large number of evaluations of \eqref{eq:RobustNMPC:Policy:Softmax} for generating each sample of $\pi_{\vect\theta}\left[ \, \vect a\,|\,\vect s\,\right]$.
%of the MHA type is that they generate samples from the desired distribution $\pi_{\vect\theta}\left[ \, \vect a\,|\,\vect s\,\right]$ via running a Markov Chain where a possibly fairly large number of 
%evaluations of the policy $\pi_{\vect\theta}\left[ \, \vect a\,|\,\vect s\,\right]$ to generate each sample of \eqref{eq:MPC:Softmax}. In the context of the softmax approach, a possibly large number of evaluations of \eqref{eq:RobustNMPC:Policy:Softmax} is then required to generate a sample of $\pi_{\vect\theta}\left[ \, \vect a\,|\,\vect s\,\right]$. 
Hence, while the simplicity of this approach is appealing, it presents the significant drawbacks that sampling policy \eqref{eq:MPC:Softmax} can be very expensive. This difficulty is arguably alleviated in the static case \eqref{eq:RobustNMPC:Policy:Constraints:Static}, where \eqref{eq:RobustNMPC:Policy:Softmax} simplifies to:
\begin{align}
\Phi^\star_\tau\left(\vect{a},\vect s,\vect{\theta}\right) = \Phi(\vect s,\vect{a},\vect{\theta}) - \tau \sum_i\log\left(\vect{h}_i\left(\vect s,\vect{a},\vect{\theta}\right)\right),
\label{eq:RobustNMPC:Policy:Softmax:Static}
\end{align}
and the evaluation of \eqref{eq:MPC:Softmax} reduces to an evaluation of the cost and constraints in \eqref{eq:RobustNMPC:Policy:Softmax:Static}. We now turn to another option for building safe policies, which is more adequate for a deployment in the NMPC case.
\subsection{Optimization-based safe stochastic policy} \label{sec:IPsampling}
We will consider a stochastic policy that generates control inputs 
\begin{align}
\vect a \sim{\pi}_{\vect{\theta}}\left[\vect a\,|\,\vect s\right] \label{eq:RobustNMPC:Disturbed:Policy}
\end{align}
computed from $\vect a =  \vect{u}_0^\mathrm{d}\left(\vect s,\vect\theta,\vect d\right)$ where $\vect{u}_0^\mathrm{d}$ is generated by the randomly disturbed NLP
\begin{subequations}
\label{eq:RobustNMPC:Disturbed}
\begin{align}
 \vect{u}^\mathrm{d}\left(\vect s,\vect\theta,\vect d\right) = \mathrm{arg}\min_{\vect{u}}&\quad\Phi^{\vect d}(\vect{x},\vect u,\vect{\theta},\vect d)\\% + \vect{d}^\top\vect u_0\\%+ \vect{d}^\top\vect{u}_0\\
\mathrm{s.t.}&\quad \vect f\left(\vect x,\vect u,\vect s,\vect{\theta}\right) = 0, \label{eq:Dynamics:Disturbed} \\
&\quad \vect{h}\left(\vect{x},\vect{u},\vect{\theta}\right) \leq 0,
\end{align}
\end{subequations}
for an arbitrary cost function $\Phi^\mathrm{d}(\vect{u},\vect s,\vect{\theta},\vect{d})$, and where the parameter $\vect{d}\in\mathbb{R}^{n_{\vect a}}$ is drawn from an arbitrary probability distribution, of density $\varrho(\vect d,\Sigma)$, which can, e.g., be a simple Gaussian distribution. One can readily observe that any realization of the inputs 
\begin{align}
\label{eq:dtoa}
\vect a =  \vect{u}_0^\mathrm{d}\left(\vect s,\vect\theta,\vect d\right)
\end{align}
stemming from \eqref{eq:RobustNMPC:Disturbed} is in $\mathbb{S}(\vect s)$ by construction. A simple choice for the cost function $\Phi^\mathrm{d}(\vect{u},\vect s,\vect{\theta},\vect{d})$ is via a gradient disturbance:
\begin{align}
\Phi^\mathrm{d}(\vect{u},\vect s,\vect{\theta},\vect{d}) &= \Phi(\vect{u},\vect s,\vect{\theta})+ \vect{d}^\top\vect u_0. \label{eq:Cost:GradientDisturbance}
\end{align}
The choice of cost \eqref{eq:Cost:GradientDisturbance} entails that the random variable $\vect d$ yields a gradient disturbance in the original problem \eqref{eq:Generic:NLP}, and introduces stochasticity in the inputs $\vect a$ generated.

One can readily observe that generating a sample from \eqref{eq:RobustNMPC:Disturbed:Policy} requires one to only generate a sample from the chosen density $\varrho(\vect d,\Sigma)$ and to solve the disturbed NMPC problem \eqref{eq:RobustNMPC:Disturbed}. It is therefore dramatically less expensive than the resampling and softmax approach of Sections \ref{sec:Resampling} and \ref{sec:IPsoftmax}. We will show next that computing the gradient of the score function of \eqref{eq:RobustNMPC:Disturbed:Policy} does not require any sampling, provided that the adequate algorithmic tools are adopted.

\section{Policy gradient for optimization-based Safe stochastic policy} \label{sec:Numerics}
\begin{figure}
	\includegraphics[width=1\linewidth,clip, trim = 60 180 65 390]{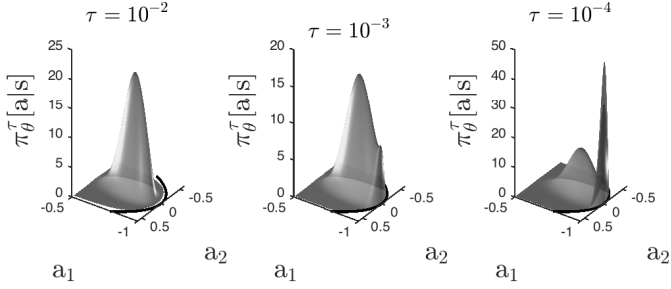} %trim= 60 300 65 300
	\caption{Illustration of the stochastic policy resulting from \eqref{eq:RobustNMPC:Disturbed:Policy}-\eqref{eq:dtoa} for different values of $\tau$ for a fixed $\vect s$, and $\vect u_0^{\mathrm d}$ restricted within a set $\mathbb{S}(\vect s)$ depicted as the solid line. The resulting probability density $\pi^\tau_{\vect\theta}\left[\vect a|\vect s\right]$ is constrained to remain within $\mathbb{S}(\vect s)$. For very low values of $\tau$, the density tends to a Dirac-like distribution on the border of the set, see right-side graph.}
	\label{fig:pivstau}
\end{figure}

We develop next the gradient of the score function associated to \eqref{eq:RobustNMPC:Disturbed:Policy}-\eqref{eq:RobustNMPC:Disturbed}. Unfortunately, a technical difficulty must be first alleviated here. Indeed, evaluating the stochastic policy ${\pi}_{\vect{\theta}}\left[\vect a\,|\,\vect s\right]$ resulting from \eqref{eq:RobustNMPC:Disturbed:Policy}-\eqref{eq:RobustNMPC:Disturbed} is in general very difficult, because it cannot be simply expressed as a function of the probability density $\varrho(\vect{d},\Sigma)$. 
%\textcolor{blue}{To support this statement, let us consider the choice of cost \eqref{eq:Cost:DirectDisturbance} with $\nu=0,\, p =2$ and assume that $\mathbb{S}\left(\vect s\right)$ is a convex polytope. Then if $\vect a^\mathrm{s}$ is inside a facet of $\mathbb{S}\left(\vect s\right)$, i.e. $\vect a^\mathrm{s}\in\partial\mathbb{S}\left(\theta\right)$, then
%\begin{align}
%\label{eq:StochPi:Facet}
%\vect{\pi}_{\vect{\theta}}\left[\,\vect a\,|\,\vect s\,\right] = \delta_{\partial\mathbb{S}\left(\vect s\right)}\left(\vect{u}\right)\int_0^\infty
%\varrho\left(\vect a + \alpha \vect{n}-\vect{\pi}_{\vect{\theta}},\Sigma\right)\mathrm{d}\alpha
%\end{align}
%where $\vect{n}$ is the exterior normal to the facet, and $\delta_{\partial\mathbb{S}\left(\vect s\right)}\left(\vect{u}\right)$ is the surface delta function \cite{} associated to the boundary $\partial\mathbb{S}\left(\vect s\right)$. For e.g. $\varrho$ Normal centered, the integral in \eqref{eq:StochPi:Facet} takes an explicit form, and \eqref{eq:StochPi:Facet} can be easily evaluated and differentiated. Unfortunately, other cases evaluating $\vect{\pi}_{\vect{\theta}}\left[\,\vect a\,|\,\vect s\,\right]$ can be very intricate.} 
The mapping $\vect d$ to $\vect u_0^{\mathrm d}$ generated by the NLP \eqref{eq:RobustNMPC:Disturbed} is in general not bijective, as it acts as a (possibly nonlinear) projection operator of the distribution $\varrho$ into the safe set $\mathbb{S}\left(\vect s\right)$. A practical outcome of  $\vect u_0^{\mathrm d}$ being non-bijective is that the resulting stochastic policy becomes Dirac-like on the boundary of the safe set $\mathbb{S}\left(\vect s\right)$, see Fig. \ref{fig:pivstau} for an illustration. %\textcolor{blue}{Here an illustration would really help}. This effect is in general difficult to handle.

%let us e.g. consider the cost \eqref{eq:Cost:DirectDisturbance} with $\alpha=0,\, p =2$. Then for $\vect a\in\mathrm{int}\left(\mathbb{A}\left(\theta\right)\right)$ the stochastic policy is trivially given by:

%\begin{align}
%\vect{\pi}_{\vect{\theta}}\left[\,\vect{u}\,|\,\vect{s}\,\right] = \frac{\delta_{\partial\mathbb{S}}\left(\vect{u}\right)}{\sqrt{2\pi\mathrm{det}( \Sigma)}}\int_0^\infty e^{-\frac{1}{2}\left(\vect{u} + \widehat{\nabla_{\vect{u}}\vect{h}}\alpha-\vect{\pi}_{\vect{\theta}}\right)^\top\Sigma^{-1}\left(\star\right)}\mathrm{d}\alpha
%\end{align}
%where $\widehat{\nabla_{\vect{u}}\vect{h}} = \frac{\nabla_{\vect{u}}\vect{h}}{\left\|\nabla_{\vect{u}}\vect{h}\right\|}$, and $\delta_{\partial\mathbb{S}}\left(\vect{u}\right)$ is the surface delta function \cite{} associated to the boundary $\partial\mathbb{S}$ of $\mathbb{S}$. 
% \textcolor{blue}{is fucked up}

In order to alleviate this difficulty, similarly to the developments of Sec. \ref{sec:IPsoftmax},  we will cast \eqref{eq:RobustNMPC:Disturbed} in an interior-point context. For computational reasons, we will consider the primal-dual interior point formulation of \eqref{eq:RobustNMPC:Disturbed}  \cite{Biegler2010}, which have the First-Order Necessary Conditions (FONC):
%\begin{align}
%\label{eq:StochasticFullInputSample}
%\vect r_\tau(\vect z,\vect \theta,\vect d) = \matr{c}{\nabla_{\vect u}\Phi^\mathrm{d} + \nabla_{\vect u}\vect h \vect \mu  \\
%\mathrm{diag}(\vect\mu) \vect h+\tau} = 0 
%\end{align}
\begin{align}
\label{eq:StochasticFullInputSample}
\vect r_\tau(\vect z,\vect \theta,\vect d) = \matr{c}{\nabla_{\vect w}\Phi^\mathrm{d} + \nabla_{\vect w}\vect h \vect \mu + \nabla_{\vect w}\vect f \vect \lambda  \\
\vect f \\
\mathrm{diag}(\vect\mu) \vect h+\tau} = 0 
\end{align}
for $\tau > 0$ and under the conditions $\vect h < 0,\quad \vect\mu > 0$. Here we label $\vect w = \left\{\vect u,\vect x\right\}$ and $\vect z = \left\{\vect w,\,\vect\lambda,\vect\mu\right\}$ the primal-dual variables of \eqref{eq:StochasticFullInputSample}. We will label $\vect{u}^{\tau}\left(\vect s,\vect{\theta},\vect{d}\right)$ the parametric primal solution of \eqref{eq:StochasticFullInputSample}, and $\pi_{\vect\theta}^\tau\left[\vect a|\vect s\right]$ the stochastic policy resulting from using $\vect a = \vect u_0^\tau\left(\vect s,\vect\theta,\vect d\right)$. Under standard regularity assumptions \cite{} on \eqref{eq:RobustNMPC:Disturbed}, the algebraic conditions \eqref{eq:StochasticFullInputSample} admit a primal-dual solution that matches the solution of \eqref{eq:RobustNMPC:Disturbed} with an accuracy at the order of the relaxation parameter $\tau$. Moreover, the solution $\vect{u}^{\tau}\left(\vect s,\vect{\theta},\vect{d}\right)$ is guaranteed to satisfy the constraints of \eqref{eq:RobustNMPC:Disturbed}, hence it delivers safe policies. Additionally, \eqref{eq:StochasticFullInputSample} is smooth and the mapping $\vect d$ to $\vect u_0^{\mathrm d}$ becomes bijective under some mild conditions.
%As a result, the NLP \eqref{eq:RobustNMPC:Disturbed} can be made bijective, and the policy evaluated using classic techniques from calculus. The primal-dual interior point method considers that the solutions of \eqref{eq:RobustNMPC:Disturbed} are obtained from solving the relaxed KKT conditions
%
%The error between the true solution of \eqref{eq:RobustNMPC:Disturbed} and the one delivered by solving \eqref{eq:StochasticFullInputSample} is of the order of the relaxation parameter $\tau$ 
We therefore propose to use \eqref{eq:StochasticFullInputSample} as a smooth surrogate for \eqref{eq:RobustNMPC:Disturbed}. We will then use the sensitivities of \eqref{eq:StochasticFullInputSample} to compute the gradient of the score function of $\pi_{\vect\theta}^\tau$.

Fig. \ref{fig:pivstau} provides an illustration of the stochastic policy delivered by \eqref{eq:StochasticFullInputSample} for different values of $\tau$, and how the stochastic policy adopts a Dirac-like shape on the border of the safety set when $\tau\rightarrow 0$.

In the following, we will use the notation $\vect{g}$ for the first block of $m$ inputs resulting from \eqref{eq:StochasticFullInputSample}, i.e.: 
\begin{align}
\label{eq:g_function}
\vect{g}\left(\vect s,\vect{\theta},\vect{d}\right)  = \vect{u}^{\tau}_0 \left(\vect s,\vect{\theta},\vect{d}\right) \approx  \vect{u}_0^\mathrm{d}\left(\vect s,\vect\theta,\vect d\right) ,
\end{align}
delivering $\vect a$ from \eqref{eq:StochasticFullInputSample}. The stochastic policy $\vect{\pi}_{\vect{\theta}}$ then results from the transformation of the probability density $\vect{d}\sim\varrho(\vect{d},\Sigma)$ via $\vect{g}$, and can be evaluated using  \cite{Bertsekas2002}
\begin{align}
\label{eq:PolicyEvaluationStochastic}
{\pi}_{\vect{\theta}}\left[\vect a\,|\,\vect s\right] =\left. \varrho\left(\vect{g}^{-1},\Sigma\right) \det\left(\frac{\partial \vect{g}^{-1}}{\partial \vect a}\right)\right|_{\vect a,{\vect{\theta}},\vect s},
\end{align}
where function $\vect{g}^{-1}$ is such that
\begin{align}
\vect{d} = \vect{g}^{-1}\left(\vect a,\vect{\theta},\vect s\right) \label{eq:InputTod}
\end{align}
for any $\vect{d}$ and associated $\vect a$ delivered by \eqref{eq:g_function}. The (local) existence of $\vect{g}^{-1}$ is guaranteed by the implicit function theorem if $\frac{\partial \vect{g}}{\partial \vect{d}}$ is full rank. We will use \eqref{eq:PolicyEvaluationStochastic} to compute the gradient of the score function of $\vect\pi_{\vect \theta}$.

For the sake of completeness, we provide hereafter a Lemma establishing the rank of $\frac{\partial \vect{g}}{\partial \vect{d}}$ for the gradient disturbance strategy \eqref{eq:Cost:GradientDisturbance}.
\begin{Lemma}\label{Lemma:dgdd:rank} 
	For the choice of cost function \eqref{eq:Cost:GradientDisturbance}, and if \eqref{eq:RobustNMPC:Disturbed} satisfies LICQ and SOSC, the Jacobian $\frac{\partial \vect{g}}{\partial \vect{d}}$ of function $\vect g$ implicitly defined by \eqref{eq:StochasticFullInputSample}-\eqref{eq:g_function} is full rank for any $\tau >0$.
\end{Lemma}

\begin{IEEEproof} for the sake of simplicity, we will prove the result using the primal interior-point conditions corresponding to \eqref{eq:StochasticFullInputSample}. The Lemma will then hold from the equivalence between the primal-dual and primal interior-point problem \cite{Nocedal2006}. The primal interior-point conditions read as \cite{Nocedal2006}:
\begin{align}
\label{eq:StochasticFullInputSample:Primal}
%\vect r^{\mathrm{p}}_\tau(\vect z,\vect \theta,\vect d) = 
\matr{c}{\nabla_{\vect w}\Phi^\mathrm{d} + \tau\nabla_{\vect w}\vect h\mathrm{diag}(\vect h)^{-1} + \nabla_{\vect w}\vect f \vect \lambda  \\
\vect f } = 0.
\end{align}
The Implicit Function Theorem (IFT) then guarantees that: 
\begin{align}
\matr{cc}{H& \nabla_{\vect w}\vect f \\
\nabla_{\vect w}\vect f^\top & 0}\matr{c}{\frac{\partial \vect w}{\partial \vect d} \\ \frac{\partial \vect \lambda}{\partial \vect d}} =- \matr{c}{\nabla_{\vect w\vect d}\Phi^\mathrm{d} \\ 0}, %= -\matr{c}{I_{m\times m} \\ 0 }
\end{align}
where $H$ is the Jacobian of the first row in \eqref{eq:StochasticFullInputSample:Primal}. Defining $\mathcal N$ the null space of $\nabla_{\vect w}\vect f^\top $, i.e., $\nabla_{\vect w}\vect f^\top\mathcal N=0$, one can verify that using $\nabla_{\vect u_0\vect d}\Phi^\mathrm{d} = I_{{n_{\vect a}}\times {n_{\vect a}}}$ from \eqref{eq:Cost:GradientDisturbance}:
\begin{align}
\frac{\partial \vect w}{\partial \vect d} & =  -\mathcal N\left(\mathcal N^\top H \mathcal N\right)^{-1}\mathcal N^\top \nabla_{\vect w\vect d}\Phi^\mathrm{d}\\
&=- \mathcal N\left(\mathcal N^\top H \mathcal N\right)^{-1}\mathcal N_0^\top, %\matr{c}{\nabla_{\vect w\vect d}\Phi^\mathrm{d} \\ 0}
\end{align}
where 
%\begin{align}
$\mathcal{N}_0 = \matr{cccc}{I_{{n_{\vect a}}\times {n_{\vect a}}} & 0 & \ldots & 0}\mathcal{N}$.
%\end{align}
 The invertibility of $\mathcal N^\top H \mathcal N$ is guaranteed if \eqref{eq:RobustNMPC:Disturbed} satisfies LICQ and SOSC. It follows that
\begin{align}
\frac{\partial \vect g}{\partial \vect d}  =- \mathcal N_0\left(\mathcal N^\top H \mathcal N\right)^{-1}\mathcal N_0^\top. \label{eq:dgdd:rank}
\end{align}
Since the dynamics $\vect f$ cannot restrict the input $\vect u$ in \eqref{eq:RobustNMPC:Disturbed}, $\mathcal N$ spans the full space of $\vect u$, and therefore $\mathcal{N}$ must span the full input space for $\vect u_0$, such that $\mathcal N_0$ is full rank. As a result, \eqref{eq:dgdd:rank} is full rank. %\textcolor{red}{Make the explaination easier, e.g. "dynamics do not block the inputs"}
\end{IEEEproof}
One can observe that Lemma \ref{Lemma:dgdd:rank} is also trivially valid in the static case. We ought to caveat Lemma \ref{Lemma:dgdd:rank} by observing that while matrix $\frac{\partial \vect g}{\partial \vect d} $ is full rank for any $\tau >0$, it can nonetheless tend to a rank-deficient matrix for $\tau\rightarrow 0$. This issue will be discussed in Proposition \ref{eq:StochPolicy:WellDefined} and in the following remarks.

The following Lemma provides the sensitivity of function \eqref{eq:InputTod}, which will be required to compute the stochastic policy gradient. 
\begin{Lemma} If \eqref{eq:RobustNMPC:Disturbed} satisfies SOSC and LICQ then the following equalities hold:%the sensitivity $\frac{\partial \vect{g}^{-1}}{\partial\vect{\theta}}$%$\nabla_{\vect{\theta}} \left(\vect{g}^{-1}\right)$ 
%is solution of
\begin{subequations}
\label{eq:g:equalities}
\begin{align}
\label{eq:ginv:sens}
\frac{\partial \vect{g}^{-1}}{\partial\vect{\theta}}  &= -\left(\frac{\partial \vect{g}}{\partial \vect{d}}\right) ^{-1}\frac{\partial \vect{g}}{\partial\vect{\theta}}, \\
\frac{\partial \vect{g}^{-1}}{\partial\vect a} &= \left(\frac{\partial\vect g}{\partial \vect d}\right)^{-1}, \label{eq:ginv:sens2}
%\nabla_{\vect{\theta}} \left(\vect{g}^{-1}\right) \nabla_{\vect{d}} \vect{g}+ \nabla_{\vect{\theta}} \vect{g} = 0  \label{eq:Sensitivity1}
\end{align}
\end{subequations}
for any $\vect a,\vect{\theta},\vect s$ and $\vect{d} =  \vect{g}^{-1}\left(\vect a,\vect{\theta},\vect s\right)$.
%where 
%\begin{align}
% \frac{\partial \vect{g}}{\partial\vect d} = -I_0^\top\nabla^2_{\vect{u}}\Phi_\tau^{-1}\nabla_{\vect{u}\vect d}\Phi_\tau ,\quad \frac{\partial \vect{g}}{\partial\vect{\theta}}  =-I_0^\top \nabla^2_{\vect{u}}\Phi_\tau^{-1}\nabla_{\vect{u}\vect{\theta}}\Phi_\tau    \label{eq:Sensitivity2}
%\end{align}
%and matrix $I_0$ reads as:
%\begin{align}
%I_0^\top = \matr{cccc}{I & 0 & \ldots & 0}
%\end{align}
\end{Lemma}
\begin{IEEEproof} We observe that
\begin{align}
\vect{g}\left(\vect s, \vect \theta, \vect g^{-1}\left(\vect a,\vect \theta,\vect s\right)\right) = \vect a,\quad \forall\, \vect a,\vect \theta, \vect s.
\end{align}
It follows that
\begin{align}
\frac{\mathrm{d}}{\mathrm{d} \vect\theta} \vect{g}\left(\vect s, \vect \theta, \vect g^{-1}\left(\vect a,\vect \theta,\vect s\right)\right) = \frac{\partial\vect g}{\partial \vect \theta} +  \frac{\partial\vect g}{\partial \vect d}\frac{\partial \vect g^{-1}}{\partial \vect\theta} = 0,
\end{align}
which establishes \eqref{eq:ginv:sens}. Moreover, we observe that
\begin{align}
\frac{\mathrm{d}}{\mathrm{d} \vect a} \vect{g}\left(\vect s, \vect \theta, \vect g^{-1}\left(\vect a,\vect \theta,\vect s\right)\right) = \frac{\partial\vect g}{\partial \vect d} \frac{\partial \vect{g}^{-1}}{\partial \vect a} = I,
\end{align}
which establishes \eqref{eq:ginv:sens2}.
%\begin{align}
%\nabla_{\vect u} \Phi_\tau\left(\vect u^\tau\left(\vect s,\vect{\theta},\vect{d}\right) ,\vect s, \vect \theta,\vect d\right) = 0,\quad \forall\,  \vect \theta,\vect d,\vect s
%\end{align}
%such that
%\begin{align}
%\frac{\partial \vect u^\tau}{\partial \vect{d}} = -\nabla^2_{\vect{u}}\Phi_\tau^{-1}\nabla_{\vect{u}\vect{d}}\Phi_\tau,\quad \frac{\partial \vect u^\tau}{\partial \vect{\theta}} = -\nabla^2_{\vect{u}}\Phi_\tau^{-1}\nabla_{\vect{u}\vect{\theta}}\Phi_\tau 
%\end{align}
%Equations \eqref{eq:Sensitivity2} can then be assembled.
\end{IEEEproof}

\subsection{Gradient of the score function}

We can then use \eqref{eq:PolicyEvaluationStochastic} to develop expressions for computing the gradient of the policy score function $\nabla_{\vect\theta}\log \vect{\pi}_{\vect{\theta}}$. This is detailed in the following Proposition.
\begin{Proposition} The gradient of the score function for a given realization of $\vect a$ obtained from a realization of $\vect d$ via solving \eqref{eq:RobustNMPC:Disturbed} reads as:
\begin{align}
&\nabla_{{\theta}}\log {\pi}_{\vect{\theta}}\left[\vect a\,|\,\vect s\right] = \label{eq:AwesomeResult} \vect m -\left(\varrho^{-1}\frac{\partial \varrho}{\partial \vect d}\left(\frac{\partial \vect g}{\partial \vect d}\right)^{-1}\frac{\partial \vect g}{\partial \vect\theta}\right)^\top,
\end{align}
evaluated at $\vect s,\vect{\theta},\vect{d}$, and where 
\begin{align}
\label{eq:AwesomeResult2}
\vect m_i = \left.\mathrm{Tr}\left(\frac{\partial \vect g}{\partial \vect d}\frac{\mathrm{d}}{\mathrm{d}\vect{\theta}_i}\frac{\partial \vect{g}^{-1}}{\partial \vect a}\right)\right|_{\vect s,\vect{\theta},\vect{d},\vect a}.
\end{align}
\end{Proposition} 
Computational techniques to evaluate \eqref{eq:AwesomeResult}-\eqref{eq:AwesomeResult2} are provided in Section \ref{sec:Sensitivities}.
\begin{IEEEproof}
Using \eqref{eq:PolicyEvaluationStochastic}, the score function of ${\pi}_{\vect{\theta}}\left[\vect a\,|\,\vect s\right]$ is given by:
\begin{align}
\label{eq:AR:FirstStep}
\log {\pi}_{\vect{\theta}}\left[\vect a\,|\,\vect s\right] = %\left.
\log \varrho\left(\vect{g}^{-1},\Sigma\right)-\log \det\left(\frac{\partial \vect{g}^{-1}}{\partial \vect a}\right). %\right|_{\vect s,{\vect{\theta}},\vect a},
\end{align}
%where $\vect{d} =  \vect{g}^{-1}\left(\vect a,\vect{\theta},\vect s\right)$ must be read as a function of $\vect{\theta}$ in \eqref{eq:AR:FirstStep}. 
Using \eqref{eq:ginv:sens} we observe that:
%\begin{subequations}
\begin{align}
\label{eq:GradientStochastic:SecondTerm}
\nabla_{\vect\theta} \log \varrho\left(\vect{g}^{-1},\Sigma\right) &= \left.\left(\varrho^{-1} \frac{\partial \varrho}{\partial \vect d} \frac{\partial \vect g^{-1}}{\partial \vect \theta}\right)^\top\right|_{\vect s,\vect{\theta},\vect{d}} \\
&=-\left.\left(\varrho^{-1}\frac{\partial \varrho}{\partial \vect d}\left(\frac{\partial \vect g}{\partial \vect d}\right)^{-1}\frac{\partial \vect g}{\partial \vect\theta}\right)^\top\right|_{\vect s,\vect{\theta},\vect{d}}, \nonumber
\end{align}
hence providing the second term in \eqref{eq:AwesomeResult}. From calculus and using \eqref{eq:ginv:sens2}, we get:
\begin{align}
\label{eq:GradientStochasticTrick}
\frac{\mathrm d}{\mathrm{d}\vect{\theta}_i}\log \det\left(\frac{\partial \vect{g}^{-1}}{\partial \vect a}\right) &= \mathrm{Tr}\left(\left(\frac{\partial \vect{g}^{-1}}{\partial \vect a}\right)^{-1} \frac{\mathrm d}{\mathrm{d}\vect{\theta}_i}\frac{\partial \vect{g}^{-1}}{\partial \vect{a}}\right) \\
&= \mathrm{Tr}\left(\frac{\partial \vect{g}}{\partial \vect{d}}\frac{\mathrm d}{\mathrm{d}\vect{\theta}_i}\frac{\partial \vect{g}^{-1}}{\partial \vect{a}}\right) = \vect m_i, \nonumber
\end{align}
hence providing \eqref{eq:AwesomeResult2} component-wise.
\end{IEEEproof}
We now turn to detailing how the sensitivities of functions $\vect g$ and $\vect g^{-1}$ can be computed at limited computational cost.
\subsection{Sensitivity computation} \label{sec:Sensitivities}
We provide hereafter some expressions allowing one to evaluate the terms in \eqref{eq:AwesomeResult}-\eqref{eq:AwesomeResult2}. First, it is useful to provide the sensitivities of function $\vect g = \vect z_0$, where $\vect z_0$ is the first $m$ elements of $\vect z$, solution of \eqref{eq:StochasticFullInputSample}. If LICQ and SOSC hold \cite{Nocedal2006} for the NLP \eqref{eq:RobustNMPC:Disturbed}, one can verify that the Implicit Function Theorem (IFT) guarantees that:
%\begin{subequations} 
\begin{align}
\label{eq:z:sensitivities}
\frac{\partial \vect r_\tau}{\partial \vect z}\frac{\partial \vect z}{\partial \vect d} + \frac{\partial \vect r_\tau}{\partial \vect d} = 0, \qquad
\frac{\partial \vect r_\tau}{\partial \vect z}\frac{\partial \vect z}{\partial \vect \theta} + \frac{\partial \vect r_\tau}{\partial \vect \theta} = 0 
\end{align}
and therefore $\frac{\partial \vect{g}}{\partial \vect{d}}$ and $\frac{\partial \vect{g}}{\partial \vect{\theta}}$, required in the second term of \eqref{eq:AwesomeResult}, can be extracted from the $m$ first rows of $\frac{\partial \vect{z}}{\partial \vect{d}}$ and $\frac{\partial \vect{z}}{\partial \vect{\theta}}$ obtained by solving the linear system \eqref{eq:z:sensitivities}. 

Obtaining the second-order term $\frac{\partial^2 \vect{g}^{-1}}{\partial \vect{\theta}_i\partial \vect{a}}$ in \eqref{eq:GradientStochasticTrick} can be fairly involved. In order to simplify its computation, we propose to use the following approach. Let us define:
\begin{align}
\tilde{\vect z} = \left\{\vect d ,\,  \vect u_1, \, \ldots, \, \vect u_{N-1}, \vect x,\, \vect\lambda, \vect\mu \right\},
\end{align}
given implicitly by \eqref{eq:StochasticFullInputSample} as a function of $\vect s, \vect \theta, \vect a$ given. One can then construe $\tilde{\vect z}$ and therefore $\vect d$ as an implicit function of $\vect s, \vect \theta, \vect u_0$, with $\vect u_0=\vect a$, defined by \eqref{eq:StochasticFullInputSample}. It follows that function $\vect g^{-1} = \tilde{\vect z}_0$, i.e., function  $\vect g^{-1}$ is given by the $m$ first entries of $\tilde{\vect z}$, implicitly defined by \eqref{eq:StochasticFullInputSample}. The IFT then naturally applies and delivers:
\begin{align}
\label{eq:z:sensitivities:reverse}
\frac{\partial \vect r_\tau}{\partial \tilde{\vect z}}\frac{\partial \tilde{\vect z}}{\partial \vect a} + \frac{\partial \vect r_\tau}{\partial \vect u_0} = 0, \qquad
\frac{\partial \vect r_\tau}{\partial \tilde{\vect z}}\frac{\partial \tilde{\vect z}}{\partial \vect \theta} + \frac{\partial \vect r_\tau}{\partial \vect \theta} = 0
\end{align}
such that $\frac{\partial \vect{g}^{-1}}{\partial \vect a}, \, \frac{\partial \vect{g}^{-1}}{\partial \vect{\theta}}$ can be extracted from the $m$ first rows of $\frac{\partial \tilde{\vect z}}{\partial \vect a},\, \frac{\partial \tilde{\vect z}}{\partial \vect \theta}$, obtained by solving  the linear system \eqref{eq:z:sensitivities:reverse}. The second-order term $\frac{\partial^2 \vect{g}^{-1}}{\partial \vect{\theta}_i\partial \vect{a}}$ in \eqref{eq:GradientStochasticTrick} can be obtained from solving the second-order sensitivity equation of the NLP:
\begin{align}
\label{eq:SecondOrderSens:Stochastic}
&\frac{\partial \vect r_\tau}{\partial \tilde{\vect z}} \frac{\partial^2 \tilde{\vect z}}{\partial \vect\theta_i \partial \vect a} +\left(\frac{\partial^2 \vect r_\tau}{\partial \vect\theta_i \partial \tilde{\vect z}} + \sum_j\frac{\partial^2 \vect r_\tau}{\partial \tilde{\vect z}\partial \vect z_j}\frac{\partial \tilde{\vect z}_j}{\partial \vect \theta_i}\right)\frac{\partial \tilde{\vect z}}{\partial \vect a} +  \frac{\partial^2 \vect r_\tau}{\partial \vect\theta_i \partial \vect a}\nonumber \\& \hspace{1cm}+ \sum_j \frac{\partial^2 \vect r_\tau}{\partial \vect a \partial \tilde{\vect z}_j}\frac{\partial \tilde{\vect z}_j}{\partial \vect \theta_i}=0.
\end{align}
One can solve the linear system \eqref{eq:SecondOrderSens:Stochastic} for $\frac{\partial^2 \tilde{\vect z}}{\partial \vect\theta_i \partial \vect a}$, and therefore obtain $\frac{\partial^2 \vect{g}^{-1}}{\partial \vect{\theta}_i\partial \vect{a}}$.

We ought to underline here that the linear systems \eqref{eq:z:sensitivities}, \eqref{eq:z:sensitivities:reverse} and \eqref{eq:SecondOrderSens:Stochastic} are large but very sparse if coming from an NMPC scheme. Their sparsity ought to be exploited for computational efficiency both when forming and solving the systems. Unfortunately, the computational complexity of evaluating the linear system \eqref{eq:SecondOrderSens:Stochastic} grows with the number of parameters $\vect\theta$, which is unfavorable for rich parametrizations of the policy.

%high-order derivatives in \eqref{eq:SecondOrderSens:Stochastic} or \eqref{eq:SecondOrderSens:Stochastic:Simplified} can be computationally fairly expensive if the sparsity of the terms is not exploited in their evaluations, especially if a large number of parameters  $\vect \theta$ is used in the NMPC scheme.

\subsection{Limit case of the gradient of the score function}
One ought to observe that matrix $\frac{\partial \vect g}{\partial \vect d}$ becomes asymptotically ($\tau\rightarrow 0$) rank deficient if some constraints are active at the first stage $k=0$, hence restricting $\vect a$ on some manifold of $\mathbb{R}^{n_{\vect a}}$. As a result, it is not obvious that the terms involved in \eqref{eq:AwesomeResult} and \eqref{eq:GradientStochasticTrick} are asymptotically well defined for $\tau \rightarrow 0$. The following proposition alleviates this concern in some cases. The other cases are discussed after the Proposition.

\begin{Proposition}\label{eq:StochPolicy:WellDefined} For the choice of cost function \eqref{eq:Cost:GradientDisturbance}, and if the MPC model dynamics and constraints are not depending on the parameters, i.e., $\nabla_{\vect\theta}\vect h=0,\,\nabla_{\vect\theta}\vect f=0 $, and if $\frac{\partial \nabla^2_{\vect w}\Phi^\mathrm{d}}{\partial \vect\theta}=0$ then the expressions 
\begin{align}
\label{eq:Prop:WellDefinedSensitivities:StochasticPolicy}
\left(\frac{\partial \vect{g}}{\partial \vect{d}}\right)^{-1} \frac{\partial \vect{g}}{\partial \vect{\theta}},\qquad \left(\frac{\partial \vect{g}}{\partial \vect{d}}\right)^{-1}\nabla_{\vect\theta_i} \frac{\partial \vect{g}}{\partial \vect{\theta}},
\end{align}
are well defined for $\tau\rightarrow 0$ if Problem \eqref{eq:RobustNMPC:Disturbed} fulfils LICQ and SOSC.
\end{Proposition}

\begin{IEEEproof}We will proceed with proving that the expressions \eqref{eq:Prop:WellDefinedSensitivities:StochasticPolicy} are well defined in the sense of the pseudo-inverse in an active-set setting deployed on \eqref{eq:RobustNMPC:Disturbed}. The asymptotic result \eqref{eq:Prop:WellDefinedSensitivities:StochasticPolicy} will then hold from the convergence of the Interior-Point solution to the active-set one. Consider $\mathbb{A}$ the (strictly) active set of \eqref{eq:RobustNMPC:Disturbed}, i.e., the set of indices $i$ such that $\vect h_i=0,\vect\mu_i >0$ at the solution. We observe that 
\begin{align}
\matr{cc}{H& \nabla_{\vect w}\vect q \\
\nabla_{\vect w}\vect q^\top & 0}\matr{c}{\frac{\partial \vect w}{\partial \vect d} \\ \frac{\partial \vect \nu}{\partial \vect d}} =- \matr{c}{\nabla_{\vect w\vect d}\Phi^\mathrm{d} \\ 0}, %= -\matr{c}{I_{m\times m} \\ 0 }
\end{align}
where $H$ is the Hessian of the Lagrange function associated to \eqref{eq:RobustNMPC:Disturbed} and
\begin{align}
\vect{q} =  \matr{c}{\vect f \\ \vect h_{\mathbb{A}}},\qquad \vect\nu=\matr{c}{\vect\lambda\\ \vect\mu_{\mathbb{A}}}.
\end{align}
Defining $\mathcal N_{\mathbb{A}}$ the null space of $\nabla_{\vect w}\vect q^\top $, i.e. $\nabla_{\vect w}\vect q^\top\mathcal N_{\mathbb{A}}=0$, and following the same line as in Lemma \ref{Lemma:dgdd:rank}, we observe that:
%one can verify that using $\nabla_{\vect u_0\vect d}\Phi^\mathrm{d} = I_{m\times m}$ from \eqref{eq:Cost:GradientDisturbance}:
%\begin{align}
%\frac{\partial \vect w}{\partial \vect d} & =  -\mathcal N\left(\mathcal N^\top H \mathcal N\right)^{-1}\mathcal N^\top \nabla_{\vect w\vect d}\Phi^\mathrm{d}\\
%&=- \mathcal N\left(\mathcal N^\top H \mathcal N\right)^{-1}\mathcal N_0^\top %\matr{c}{\nabla_{\vect w\vect d}\Phi^\mathrm{d} \\ 0}
%\end{align}
%where 
%%\begin{align}
%$\mathcal{N}_0 = \matr{cccc}{I_{m\times m} & 0 & \ldots & 0}\mathcal{N}$.
%%\end{align}
% The invertibility of $\mathcal N^\top H \mathcal N$ is guaranteed if \eqref{eq:RobustNMPC:Disturbed} satisfies LICQ and SOSC. It follows that
\begin{align}
\frac{\partial \vect g}{\partial \vect d}  =- \mathcal N_{\mathbb{A}_0}\left(\mathcal N_{\mathbb{A}}^\top H \mathcal N_{\mathbb{A}}\right)^{-1}\mathcal N_{\mathbb{A}_0}^\top ,
\end{align}
where $\mathcal{N}_{\mathbb{A}_0} = \matr{cccc}{I_{m\times m} & 0 & \ldots & 0}\mathcal{N}_{\mathbb{A}}$.
Using a similar reasoning, since $\frac{\partial \nabla_{\vect w}\vect q}{\partial \vect \theta}=0$ and $\frac{\partial H}{\partial \vect \theta}=0$, we observe that: 
\begin{align}
\partial^k_{\vect\theta} \vect g = - \mathcal N_{\mathbb{A}_0}\left(\mathcal N_{\mathbb{A}}^\top H \mathcal N_{\mathbb{A}}\right)^{-1}\mathcal N_{\mathbb{A}}^\top \partial^k_{\vect\theta} \nabla_{\vect w}\Phi^\mathrm{d},
\end{align}
where $\partial^k_{\vect\theta}$ are multi-indexed differential operators with respect to $\vect\theta$, using any multi index $k$. %\begin{align}
This entails that $\partial_{\vect\theta}^k \vect g$ can be expressed as $\partial_{\vect\theta}^k \vect g = \mathcal{N}_{\mathbb{A}_0}Q$ for some matrix $Q$. Consider then the linear system in the unknown matrix $X$: 
\begin{align}
\label{eq:LinearSystemSensitivities:General}
\frac{\partial \vect{g}}{ \partial \vect d} X + \partial_{\vect\theta}^k \vect g = 0.
\end{align}
%which equivalently reads as:
%\begin{align}
%\mathcal{N}_0\left(\mathcal{N}^\top H \mathcal{N}\right)^{-1}\mathcal{N}_0^\top X + \mathcal{N}_0 Q = 0.
%\end{align}
Since $\mathcal{N}_{\mathbb{A}}^\top H \mathcal{N}_{\mathbb{A}}$ is full rank, $ \mathcal{N}_{\mathbb{A}_0} Q $ is in the span of matrix $\mathcal{N}_{\mathbb{A}_0}\left(\mathcal{N}_{\mathbb{A}}^\top H \mathcal{N}_{\mathbb{A}}\right)^{-1}\mathcal{N}_{\mathbb{A}_0}^\top$. It follows that \eqref{eq:LinearSystemSensitivities:General} is consistent, such that it can be solved for $X$ using, e.g., a pseudo-inverse. As a result, by continuity of the solution manifold defined by \eqref{eq:StochasticFullInputSample}, the expressions \eqref{eq:Prop:WellDefinedSensitivities:StochasticPolicy} have a well-defined limit for $\tau \rightarrow 0$, given by the solution of the linear system \eqref{eq:LinearSystemSensitivities:General}.
\end{IEEEproof}

It is important to note here that Proposition \ref{eq:StochPolicy:WellDefined} relies on the safety constraints $\vect h$ being independent of the parameters $\vect \theta$. This requirement is not an artificial effect of the approach, but rather a fundamental limitation of deploying the stochastic policy gradient approach on safety sets. Indeed, one can observe that the probability density $\vect \pi_{\vect\theta}[\vect a\,|\, \vect s]$ can be discontinuous at the border $\partial\mathbb{S}\left(\vect s\right)$ of the set set $\mathbb{S}\left(\vect s\right)$ defined by $\vect h$, as the probability density is (possibly) non-zero at $\partial\mathbb{S}\left(\vect s\right)$ and zero outside. As a result, the gradient $\nabla_{{\theta}}\log {\pi}_{\vect{\theta}}\left[\vect a\,|\,\vect s\right]$ can be ill-defined for $\vect a\in \partial\mathbb{S}\left(\vect s\right)$ if the changes in the parameters $\vect\theta$ can move the set border $\partial\mathbb{S}\left(\vect s\right)$. The Interior-Point approach proposed in Section \ref{sec:IPsampling} alleviates this problem, at the expense of keeping $\tau$ finite rather than using $\tau \rightarrow 0$. The the problem is avoided by smoothing the transition from a non-zero density in $\mathbb{S}(\vect s)$ to a zero density outside. These observations are illustrated in Figures \ref{fig:pivstau}-\ref{eq:StochPolicy:WellDefined}.

\begin{figure}
	\includegraphics[width=1\linewidth,clip,trim= 60 175 65 175]{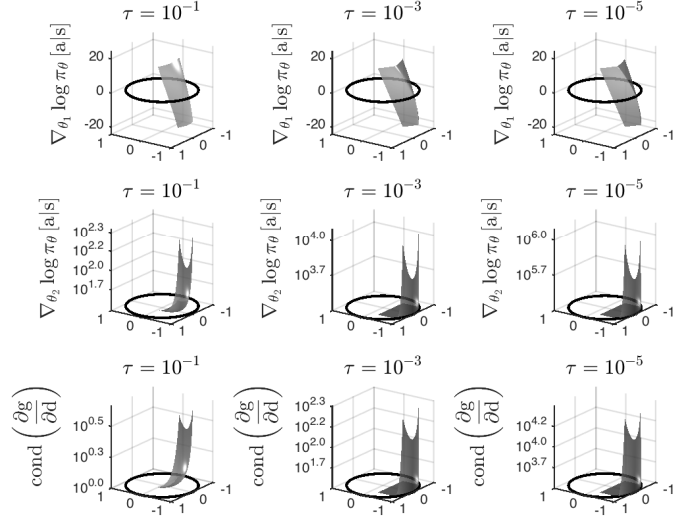} %\,
	\caption{Illustration of the gradient of the stochastic policy resulting from \eqref{eq:RobustNMPC:Disturbed:Policy}-\eqref{eq:dtoa} for different values of $\tau$, $\vect s$ fixed, and $\vect u_0^{\mathrm d}$ restricted within a set $\mathbb{S}(\vect s)$ depicted as the solid circle. The first row of graphs depict the gradient with respect to parameter $\vect\theta_1$ for which $\nabla_{\vect\theta_1}\vect h,\,\nabla_{\vect\theta_1}\vect f=0$, while the second row depicts the gradient with respect to parameter $\vect\theta_2$ for which $\nabla_{\vect\theta_2}\vect h\neq0$. The last row depicts the conditioning of matrix $\frac{\partial\vect g}{\partial \vect d}$. As predicted by Proposition \ref{eq:StochPolicy:WellDefined} for $\tau\rightarrow 0$, $\frac{\partial\vect g}{\partial \vect d}$ tends to a rank-deficient matrix for $\vect a = \vect u_0^{\mathrm d}\rightarrow \partial\mathbb{S}\left(\vect s\right)$, and the gradient $\nabla_{\vect \theta_2}\log \vect \pi_{\vect\theta}$ degenerates while $\nabla_{\vect \theta_1}\log \vect \pi_{\vect\theta}$ does not.}
	\label{fig:dpivstau}
\end{figure}

\section{Safe RL steps} \label{sec:SafeRLSteps}

The methodology described so far allows one to deploy a safe policy and safe exploration using a robust NMPC scheme, in order to compute the deterministic policy gradient, and determine directions in the parameter space $\vect \theta$ that improve the closed-loop performance of the resulting control policy. However, taking a step in $\vect \theta$ can arguably jeopardise the safety of the policy, e.g., by modifying the constraints, or the models underlying the robust NMPC scheme. The problem of modifying the NMPC parameters while maintaining safety is arguably a complex one, and beyond the scope of this paper. However, in the robust linear MPC context detailed in Section \ref{sec:SafeTree}, there is a simple approach to handle this problem, which we detail here. We observe that a classic gradient step of step-size $\alpha>0$ reads as:
\begin{align}
\label{eq:RL:GradientStep}
\vect\theta = \vect\theta_- - \alpha\nabla_{\vect\theta}J
\end{align}
where $\vect\theta_-$ is the previous vector of parameters. One can trivially observe that the gradient step can be construed as the solution of the optimization problem:
\begin{align}
\min_{\vect\theta}&\quad \frac{1}{2}\left\|\vect\theta-\vect\theta_-\right\|^2+ \alpha\nabla_{\vect\theta}J^\top\left(\vect\theta-\vect\theta_-\right).
\end{align}
Imposing the data-driven safe-design constraints \eqref{eq:DispersionApprox:Data} on the gradient step generating the new parameters can then simply be cast as the following constrained optimization problem:
\begin{subequations}
\label{eq:Constrained:RL}
\begin{align}
\min_{\vect\theta,\vect\vartheta}&\quad \frac{1}{2}\left\|\vect\theta-\vect\theta_-\right\|^2+ \alpha\nabla_{\vect\theta}J^\top\left(\vect\theta-\vect\theta_-\right) \\
\mathrm{s.t.}&\quad \vect s_{k+1} - \vect F_0\left(\vect s_k,\vect a_k,\vect \theta \right) - \sum_{i=1}^V\sum_{k=0}^{N_\mathrm{D}} \vect\vartheta_{i,k} \vect W^{i}=0, \label{eq:Inclusion1}\\
&\quad \sum_{i=1}^V \vect\vartheta_{i,k} = 1,\quad \forall k = 0,\ldots, N_{\mathcal{D}}, \\
&\quad  \vect\vartheta_{i,k} \geq 0\quad  \forall k = 0,\ldots, N_{\mathcal{D}},\quad i=1,\ldots,V, \label{eq:Inclusion3}
\end{align}
\end{subequations}
where \eqref{eq:Inclusion1}-\eqref{eq:Inclusion3} are the algebraic conditions testing \eqref{eq:DispersionApprox:Data}. We observe that unfortunately the complexity of \eqref{eq:Constrained:RL} grows with the amount of data $N_{\mathcal{D}}$ in use. In practice, the data set $\mathcal D$ should arguably be limited to incorparate relevant state transitions. A data compression technique has been proposed in \cite{Zanon2019b} to alleviate this issue in the case the nominal model $\vect F_0$ is fixed. Future work will improve on this baseline.

\section{Implementation \& illustrative Example} \label{sec:Simulations}
In this section, we provide some details on how the principle presented in this paper can be implemented, and provide an illustrative example of this implementation. At each time instant $k$, for a given state $\vect s_k$, a sample is drawn from the stochastic policy $\pi_{\vect\theta}$, computed according to \eqref{eq:StochasticFullInputSample} with $\vect d\sim\varrho(.,\Sigma)$. The gradient of the score function is then computed using \eqref{eq:AwesomeResult}. The data are collected to compute \eqref{eq:StochasticPiGradient}-\eqref{eq:deltaV} either on-the-fly or in a batch fashion. The policy gradient estimation \eqref{eq:StochasticPiGradient} is then used to compute the safe parameter update according to \eqref{eq:Constrained:RL}.
\subsection{RL approach}
In this example, the policy gradient was evaluated using batch Least-Squares Temporal-Difference (LSTD) techniques, whereby for each evaluation, the closed-loop system is run $S$ times for $N_t$ time steps, hence generating $S$ trajectory samples of duration $N_t$. The value function estimations is constructed using:% LSTD techniques, i.e. the value function estimation is given by:
\begin{subequations}
\label{eq:LSTDV}
\begin{align}
&\sum_{k=0}^{N_t}\sum_{i= 1}^S \delta^V(\vect s_{k,i},\vect a_{k,i},\vect s_{k+1,i}) \nabla_{\vect v}\hat V^{\vect v}_{\vect{\pi}_{\vect{\theta}}}\left(\vect s_{k,i}\right) = 0,  \\
&\delta^V := L(\vect s_{k,i},\vect a_{k,i})+ \gamma \hat V_{{\pi}_{\vect{\theta}}}^{\vect v}\left(\vect s_{k+1,i}\right) - \hat V_{{\pi}_{\vect{\theta}}}^{\vect v}\left(\vect s_{k,i}\right),
\end{align}
\end{subequations}
using a linear value function approximation 
\begin{align}
\hat V^{\vect v}\left(\vect s\right) =\vect\varrho \left(\vect s\right)^\top\vect v. \label{eq:Vapprox}
\end{align}
In this example, \eqref{eq:Vapprox} uses a simple quadratic function in $\vect\varrho \left(\vect s\right)$ to parametrize $\hat V^{\vect v}$.

%and using the $\vect v$ obtained from  \eqref{eq:LSTDV}, the advantage function estimation is given by:
%\begin{subequations}
%\label{eq:LSTDQ}
%\begin{align}
%&\sum_{k=0}^{N_t}\sum_{i= 1}^S \delta^V(\vect s_{k,i},\vect a_{k,i},\vect s_{k+1,i}) \nabla_{\vect w}\hat Q_{\vect{\pi}_{\vect{\theta}}}^{\vect w}\left(\vect s_{k,i},\vect a_{k,i}\right) = 0  \\
%&\delta^V := L(\vect s_{k,i},\vect a_{k,i})+ \gamma \hat V_{{\pi}_{\vect{\theta}}}^{\vect v}\left(\vect s_{k+1,i}\right) - \hat V_{{\pi}_{\vect{\theta}}}^{\vect w}\left(\vect s_{k,i}\right) % \\
%%&\hat Q_{\vect{\pi}_{\vect{\theta}}}^{\vect w}\left(\vect s_{k,i},\vect a_{k,i}\right) = \hat V_{{\pi}_{\vect{\theta}}}^{\vect v}\left(\vect s_{k,i}\right) + \hat A_{\vect{\pi}_{\vect{\theta}}}^{\vect w}\left(\vect s_{k,i},\vect a_{k,i}\right)  
%\end{align}
%\end{subequations}
%where $\hat A^{\vect w}_{\vect{\pi}_{\vect{\theta}}}$ reads as \cite{Sutton}
%\begin{align}
%\hat A^{\vect w}_{\vect{\pi}_{\vect{\theta}}}\left(\vect a,\vect s\right) = \vect w^\top \nabla_{\vect\theta}\log \vect \pi_{\vect\theta}\left[\vect a|\vect s\right] 
%\end{align}
%\textcolor{blue}{May switch to $\delta^V$ approach...}

We observe that \eqref{eq:LSTDV} is linear in the parameters $\vect v$, and therefore straightforward to solve. However, it can be ill-posed on some data sets, and ought to be solve using, e.g., a Moore-Penrose pseudo-inverse. The policy gradient estimation is then obtained using \eqref{eq:StochasticPiGradient}:
\begin{align}
\label{eq:DetPiGradient:Approx:FromData}
\widehat{\nabla_{{\theta}}\, J({\pi}_{\vect{\theta}})} =\sum_{k=0}^{N_t}\sum_{i= 1}^S \nabla_{{\theta}}\log \vect{\pi}_{\vect{\theta}}\left(\vect s_{k,i}\right) \delta^V.%\nabla_{\vect{\theta}} \vect{\pi}_{\vect{\theta}}\left(\vect s_{k,i}\right)^\top \vect w
\end{align}
\subsection{{Robust linear MPC scheme}}
While the proposed theory is not limited to linear problems, for the sake of clarity, we propose to use a fairly simple robust linear MPC example using multiple models and process noise. We will consider the policy as delivered by the following robust MPC scheme based on multiple models and a linear feedback policy:
\begin{subequations}
\label{eq:RobustMPC:ForExamples}
\begin{align}
\min_{\vect u,\vect x}&\,\, \sum_{j=0}^{N_\mathrm{M}}\left(\left\|\vect x_{j,N} - \bar{\vect x}\right\|^2%^\top  \left(\vect x_{N+1,i} - \bar{\vect x}_j\right) \\
 +\sum_{k=0}^{N-1}\left\|\matr{c}{\vect x_{j,k} - \bar{\vect x}\\ \vect u_{j,k} - \bar{\vect u}}\right\|^2\right)\\%^\top W \matr{c}{\vect x_{j,i} - \bar{\vect x}_j \\ \vect u_{j,i} - \bar{\vect u}_j}\nonumber\\
\mathrm{s.t.}&\,\, \vect x_{j,k+1} = A_0 \vect x_{j,k} + B_0 \vect u_{j,k} + \vect b_0 +  \vect W^j, \label{eq:StateDynamics}\\
&\,\, \|\vect x_{j,k}\|^2 \leq 1,\quad \forall\, j= 0,\ldots,N_\mathrm{M},\, k= 1,\ldots N,   \label{eq:StateConst}\\
&\,\, \vect x_{j,0} = \vect s,\quad \forall \, j = 1,\ldots, N_\mathrm{M}, \\
&\,\, \vect u_{j,0} = \vect u_{k,0},\quad \forall \, k,j = 0,\ldots, N_\mathrm{M},\\
&\,\, \vect u_{j,k} = \vect u_{0,k} - K\left(\vect x_{j,k} - \vect x_{0,k}\right),\,\,\, j=1,\ldots,N_\mathrm{M},
\end{align}
\end{subequations}
where $A_0$, $B_0$, $\vect b_0$ yield the MPC nominal model corresponding to $\vect F_0$, with $\vect W^0=0$, and $\vect W^{1,\ldots M}$ capture the vertices of the dispersion set outer approximation. Hence model $j=0$ serves as nominal model and models $j=1,\ldots,N_\mathrm{M}$ capture the state dispersion over time. The linear feedback matrix $K$ is possibly part of the MPC parameters $\vect\theta$, and is a (rudimentary) structure providing a feedback $\vect\pi^\mathrm{s}$ as described in Section \ref{sec:SafeSet}. In practice, \eqref{eq:RobustMPC:ForExamples} is equivalent to a tube-based MPC.

\subsection{Simulation setup \& results}
% One can observe that \eqref{eq:Input:Feedback} implements a robust MPC scheme with a (rudimentary) feedback policy $\vect \pi^\mathrm{s}$ as described in Section \ref{sec:SafeSet}.% The same cases as in Example 1 have been simulated.
%
%The feedback intrinsically delivered by the MPC scheme is represented here by using multiple input profiles beyond the first stage. The proposed scheme can be construed as a rudimentary scenario tree approach of depth one. We have selected here as safe constraint a quadratic pure state constraint \eqref{eq:StateConst}, which must be respected for all time and all possible model realisations.
%
%The example consists in the robust MPC scheme \eqref{eq:RobustMPC:ForExamples} using $M=5$ models, where the first model serves as a nominal model using
%\begin{align}
%\vect x_{0,i+1} = A_0\vect x_{0,i} + B_0\vect u_{0,i} + \vect b_0
%\end{align}
%subject to the free input profile $\vect u_{0,i}$ with $i=0,\ldots N-1$, and the other four models carry the different vertices of the dispersion set outer approximation $\vect W$, i.e.
%\begin{align}
%\vect x_{j,i+1} = A_0\vect x_{j,i} + B_0\vect u_{j,i} + \vect W^j,\quad j=1,\ldots,4
%\end{align}
%subject to the restricted input profile 
%\begin{align}
%\label{eq:Input:Feedback}
%\vect u_{j,i} = \vect u_{0,i} - K\left(\vect x_{j,i} - \vect x_{0,i}\right),\quad j=1,\ldots,4
%\end{align}
%where 
The simulations proposed here use the same setup as the companion paper \cite{Gros2019b} treating the stochastic policy gradient case, so as to make comparisons straightforward. The experimental parameters are summarized in Table \ref{tab:table1} and:
\begin{align}
\vect x_{k+1} = A_\mathrm{real} \vect x_{k} + B_\mathrm{real} \vect u_{k} + \vect n,
\end{align}
where the process noise $\vect n$ is selected Normal centred, and clipped to a ball. The real system was selected as:
\begin{align}
A_\mathrm{real} &= \kappa\matr{cc}{\cos \beta & \sin \beta\\ 
\sin \beta & \cos \beta},\,  B_\mathrm{real} = \matr{cc}{1.1 & 0 \\ 0 & 0.9}.
\end{align}
The real process noise $\vect n$ is chosen normal centred of covariance $\frac{1}{3}10^{-2} I$, and restricted to a ball of radius $\frac{1}{2}10^{-2}$. The initial nominal MPC model is chosen as:
\begin{align}
A_0 &= \matr{cc}{\cos \hat\beta & \sin \hat\beta\\ 
\sin \hat\beta & \cos \hat\beta},\,  B_0= \matr{cc}{1 & 0 \\ 0 & 1},\, \vect b_0 = \matr{c}{0\\0}.
\end{align}
and $N_\mathrm{M} = 4$ with:
\begin{subequations}
\begin{align}
\vect W^1 = \frac{1}{10}\matr{c}{-1 \\ -1},\quad \vect W^2 = \frac{1}{10}\matr{c}{+1 \\ -1} \\
\vect W^3 = \frac{1}{10}\matr{c}{+1 \\ +1},\quad \vect W^4 = \frac{1}{10}\matr{c}{-1 \\ +1}.
\end{align}
\end{subequations}
\begin{table}[h!]
  \begin{center}
    \caption{\footnotesize Simulation parameters}
    \label{tab:table1}
    \begin{tabular}{l|c|c} % <-- Alignments: 1st column left, 2nd middle and 3rd right, with vertical lines in between
   Parameter & Value & Description\\
      %$\sigma$ & $\beta$ & $\gamma$ \\
      \hline
      $\gamma$ & 0.99 & Discount factor\\
      %$\sigma$ & ... & Step size\\
      $\Sigma$ & $I$ & Exploration shape\\
      $\sigma$ & $10^{-3}$ & Exploration covariance\\
      $\tau$ & $10^{-2}$ & Relaxation parameter\\
      $\beta$ & $22^\circ$ & Real system parameter\\
            $\hat\beta$ & $20^\circ$ & Model parameter\\
      $N_t$ & $20$ & Sample length\\
       $S$ & $30$ & Number of sample per batch\\       
%       $n_{\vect a}$ & $4$ & Number of models in \eqref{eq:StateDynamics} \\
              $N$ & $10$ & MPC prediction horizon 
    \end{tabular}
  \end{center}
\end{table}

The baseline stage cost is selected as:
\begin{align}
L = \frac{1}{20}\left\|\vect x - {\vect x}_\mathrm{ref}\right\|^2 + \frac{1}{2}\left\|\vect u - {\vect u}_\mathrm{ref}\right\|^2
\end{align}
and serves as the baseline performance criterion to evaluate the closed-loop performance of the MPC scheme. 

\begin{figure}
\center
	\includegraphics[width=0.8\linewidth,clip,trim= 35 160 50 325]{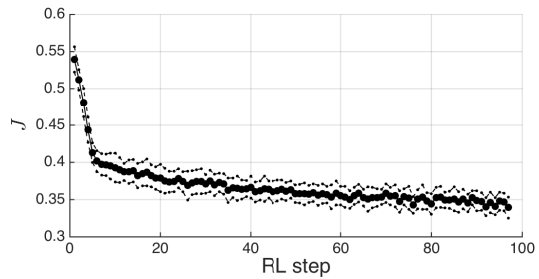} %\
	\caption{Case 1. Evolution of the closed-loop performance $J$ over the RL steps. The solid line represents the estimation of $J$ based on the samples obtained in the batch. The dashed line represent the standard deviation due to the stochasticity of the system dynamics and policy disturbances. }
	\label{fig:J1}
\end{figure}

\begin{figure}
\center
	\includegraphics[width=0.8\linewidth,clip,trim= 0 150 0 190]{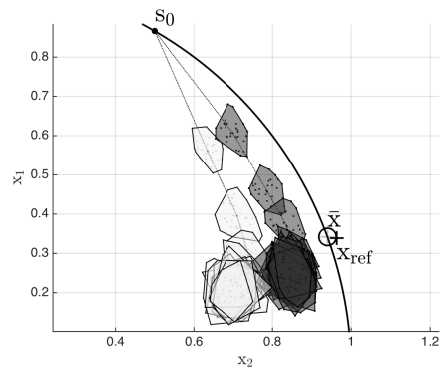} %\,trim= 0 150 0 120
	\caption{Case 1. Closed-loop system trajectories. The initial conditions $\vect s_0$ are reported, as well as the target state reference ${\vect x}_\mathrm{ref}$ (circle), and the MPC reference $\bar{\vect x}$ at the first RL step and at the last one (grey and black $+$ symbol respectively). The trajectories at the first and last RL steps are reported as the light and dark grey polytopes. The solid black curve represents the state constraint $\|\vect x\|^2 \leq 1$.}
	\label{fig:trajX1}
\end{figure}

\begin{figure}
\center
	\includegraphics[width=1\linewidth,clip,trim= 50 180 50 390]{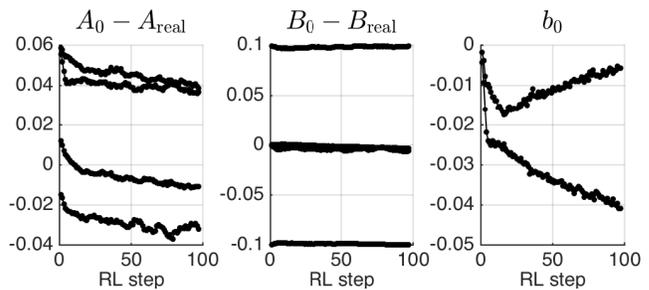} %\,
	\caption{Case 1. Evolution of the nominal MPC model over the RL steps. We report here the difference between the nominal model used in the MPC scheme and the real system.}
	\label{fig:Model1}
\end{figure}

\begin{figure}
\center
	\includegraphics[width=0.8\linewidth,clip,trim= 0 150 0 185]{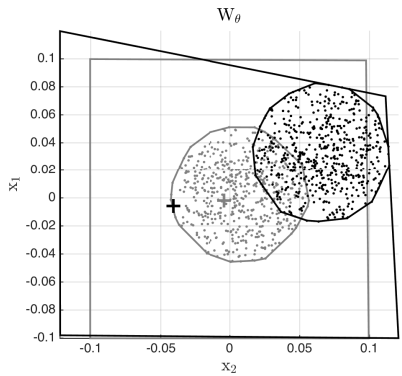} %\,trim= 0 150 0 120
	\caption{Case 1. Evolution of the MPC model biases $\vect W^{1,\ldots M}$ over the RL steps. The light grey polytope depicts the biases at the first RL step. and the points show $\vect s_{k+1} - \vect F_0\left(\vect s_k,\vect a_k,\vect \theta\right) $ for all the samples of the first batch of data. The cloud of point is inside the black thick quadrilateral thanks to the constrained RL step \eqref{eq:Constrained:RL}.  }
	\label{fig:Biases1}
\end{figure}

\begin{figure}
\center
	\includegraphics[width=0.8\linewidth,clip,trim= 0 150 0 185]{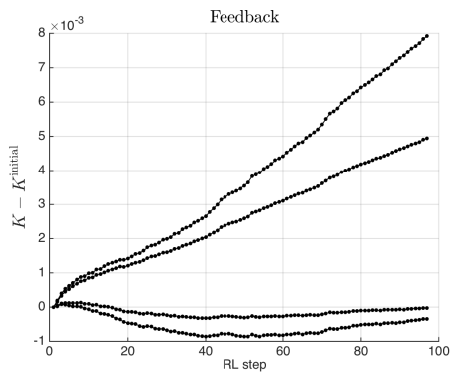} %\,trim= 0 150 0 120
	\caption{Case 1. Evolution of the MPC feedback matrix $K$ from its initial value. The feedback is only marginally adjusted by the RL algorithm. After 100 RL steps, the adaptation of the feedback gain $K$ has not yet reached its steady-state value.}
	\label{fig:Feedback1}
\end{figure}

We considered two cases, using deterministic initial conditions $\vect s_0=\matr{cc}{\cos 60^\circ&\sin 60^\circ}^\top$. Both cases consider the parameters $\vect\theta=\left\{\bar{\vect x},\,\bar{\vect u},\, A_0,\, B_0,\,\vect b_0,\, K,\, \vect W\right\}$. The first case considers a stable real system with $\kappa = 0.95$, the second case considers an unstable real system with $\kappa = 1.05$. In both cases, the target reference $\bar{\vect x}$ was provided, together with the input reference $\bar{\vect u}$ delivering a steady-state for the nominal MPC model. The feedback matrix $K$ was chosen as the LQR controller associated to the MPC nominal model. Table \ref{tab:table1} reports the algorithmic parameters. Case 1 used a step size $\alpha= 0.05$, the second case used a step size $\alpha = 0.01$. The results for the first case are reported in Figures \ref{fig:J1}-\ref{fig:Feedback1}. One can observe in Fig. \ref{fig:J1} that the closed-loop performance is improving over the RL steps. Figure \ref{fig:trajX1} shows that the improvement takes place via driving the closed-loop trajectories of the real system closer to the reference, without jeopardising the system safety. Figure \ref{fig:Model1} shows how the RL algorithm uses the MPC nominal model to improve the closed-loop performance. One can readily see from Figure \ref{fig:Model1} that RL is not simply performing system identification, as the nominal MPC model developed by the RL algorithm does not tend to the real system dynamics. %Fig. \ref{fig:Cost1} shows how the input reference is tuned by the RL algorithm. 
Figure \ref{fig:Biases1} shows how the RL algorithm reshapes the dispersion set. The upper-left corner of the set is the most critical in terms of performance, as it activates the state constraint $\|\vect x\|^2\leq 1$, and is moved inward to gain performance. The constrained RL step \eqref{eq:Constrained:RL} ensures that the RL algorithm cannot jeopardize the system safety. In Figure \ref{fig:Feedback1}, one can see that the RL algorithm does not use much the degrees of freedom provided by adapting the MPC feedback matrix $K$.

%\mario{Maybe the figures for the second case can be put next to the other ones? Or do they become too small? 4 and 9 can be put together with a different color. 6 and 11 one above the other with the same caption. 8, 13 can be superposed using different colors. 5 and 10 next to each other are too small? Then maybe one on top of the other?}
%\seb{I can look at that once the draft is more final.}

\begin{figure}
\center
	\includegraphics[width=0.8\linewidth,clip,trim= 35 160 50 325]{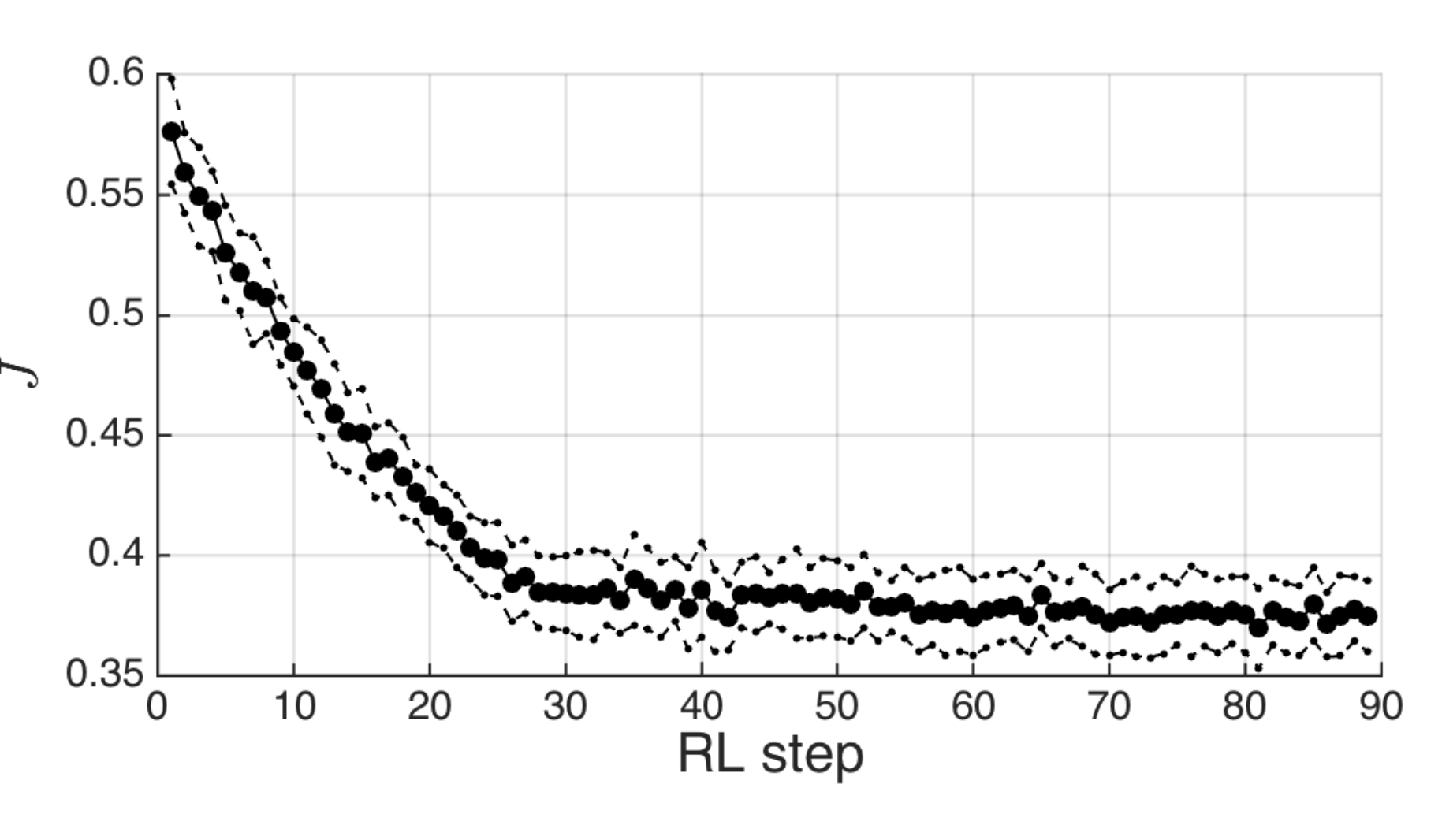} %\
	\caption{Case 2, similar to Fig. \ref{fig:J1}.}
	\label{fig:J2}
\end{figure}

\begin{figure}
\center
	\includegraphics[width=0.8\linewidth,clip,trim= 0 150 0 180]{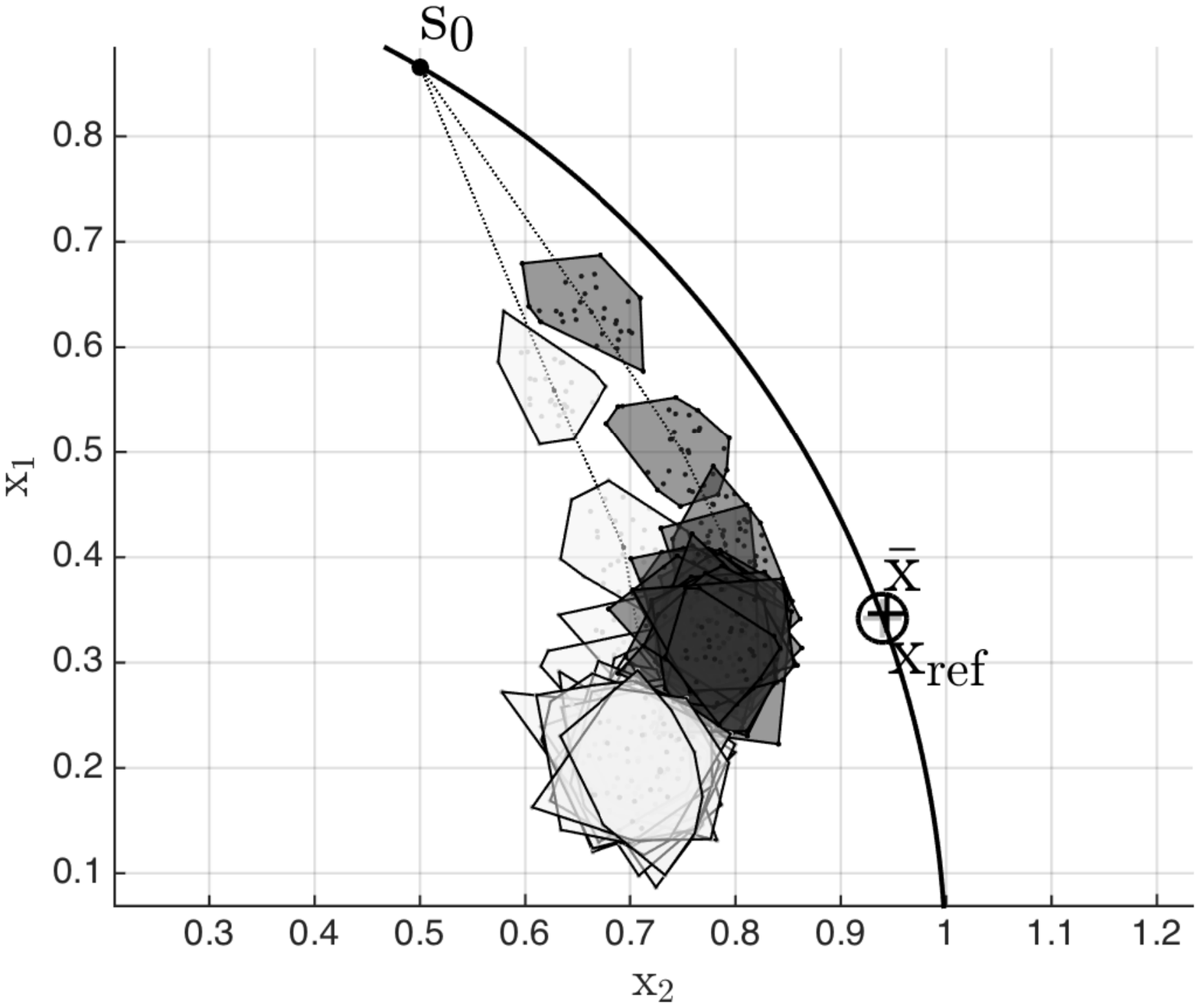} %\,trim= 0 150 0 120
	\caption{Case 2, similar to Fig. \ref{fig:trajX1}}
	\label{fig:trajX2}
\end{figure}

\begin{figure}
\center
	\includegraphics[width=1\linewidth,clip,trim= 50 180 50 390]{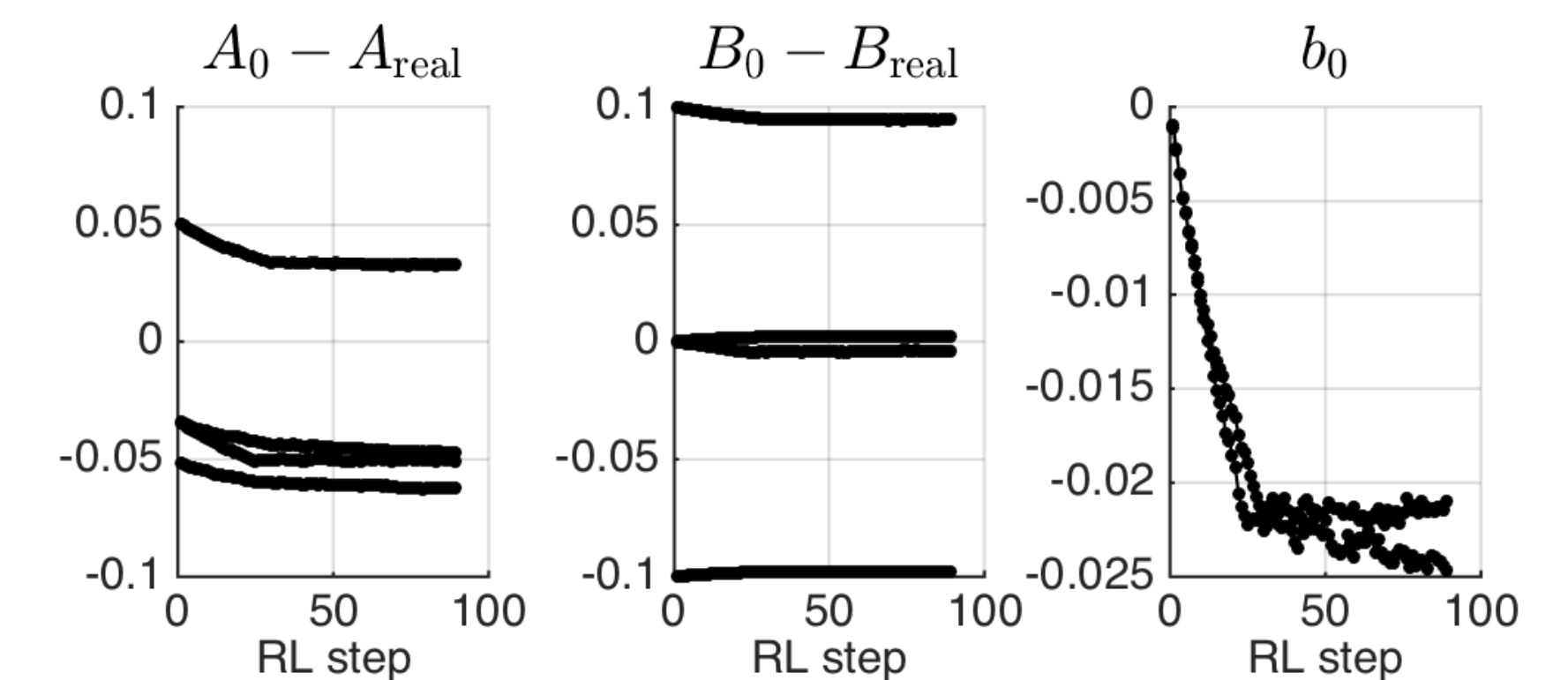} %\,
	\caption{Case 2, similar to \ref{fig:Model1}.}
	\label{fig:Model2}
\end{figure}

\begin{figure}
\center
	\includegraphics[width=0.8\linewidth,clip,trim= 0 150 0 180]{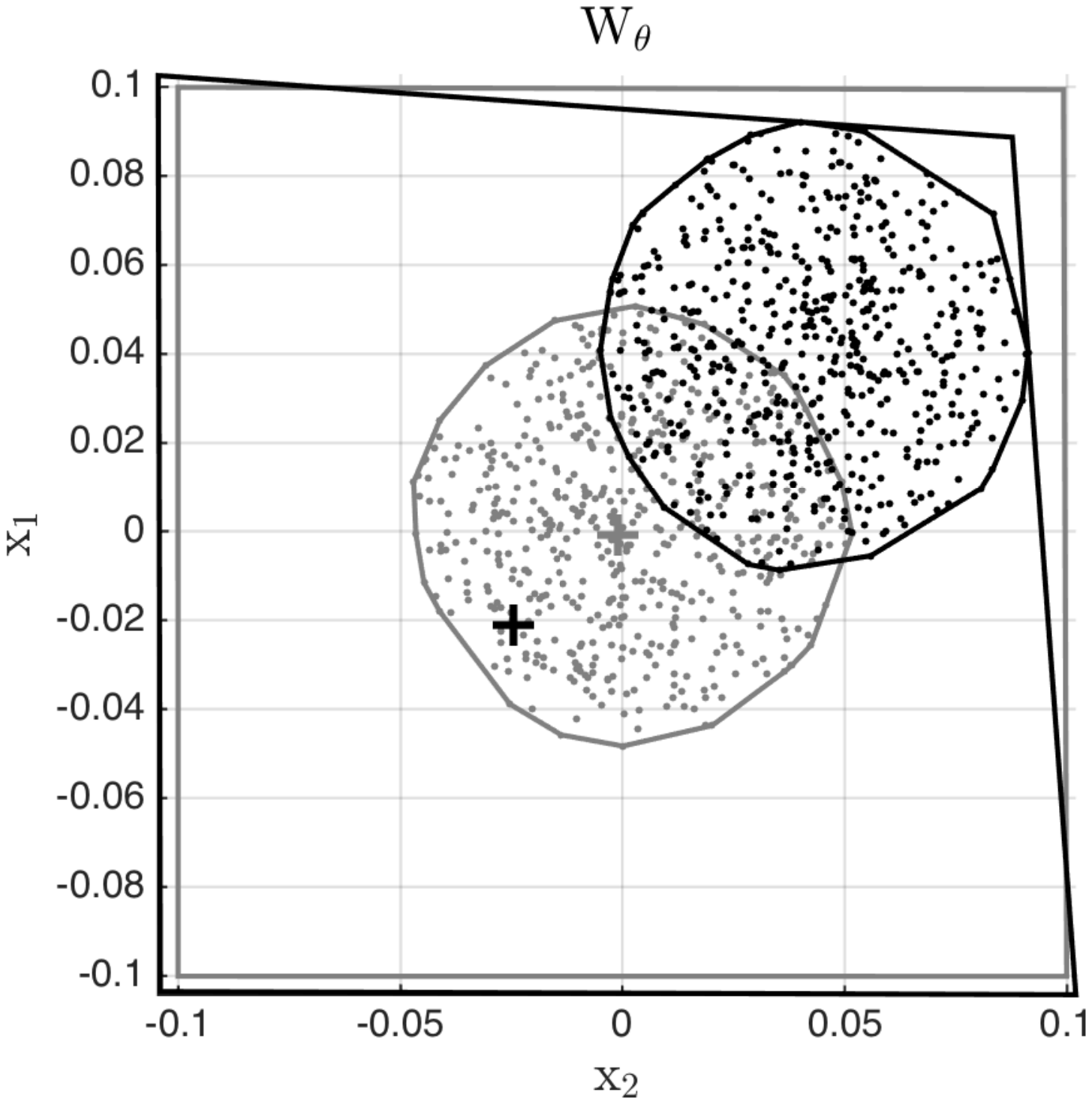} %\,trim= 0 150 0 120
	\caption{Case 2, similar to Fig. \ref{fig:Biases1}. }
	\label{fig:Biases2}
\end{figure}

\begin{figure}
\center
	\includegraphics[width=0.8\linewidth,clip,trim= 0 150 0 185]{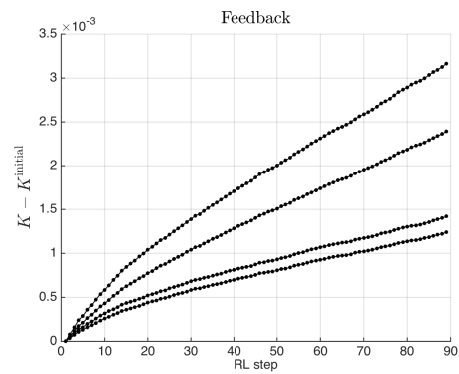} %\,trim= 0 150 0 120
	\caption{Case 2, similar to Fig. \ref{fig:Feedback1}.}
	\label{fig:Feedback2}
\end{figure}

 The results for case 2 are reported in Figures  \ref{fig:J2}-\ref{fig:Feedback2}. Similar comments hold for case 2 as for case 1. The instability of the real system does not challenge the proposed algorithm, even though a smaller step size $\sigma$ had to be used as the RL algorithm appears to more sensitive to noise. %A larger number of RL steps was used to reach a near-stationarity of the RL algorithm.

%\begin{figure}
%\center
%	\includegraphics[width=1\linewidth,clip,trim= 0 160 0 180]{Cost1} %\,
%	\caption{Case 1. Evolution of the MPC input reference $\bar{\vect u}$ and the target reference $\vect u_\mathrm{ref}$ over the RL steps. }
%	\label{fig:Cost1}
%\end{figure}

%%%% Experiment 3 %%%%
%\clearpage

%\begin{figure}
%\center
%	\includegraphics[width=1\linewidth,clip,trim= 0 160 0 180]{Cost3} %\,
%	\caption{Case 2, similar to Fig. \ref{fig:Cost1}}
%	\label{fig:Cost2}
%\end{figure}

\section{Conclusion}
This paper proposed a technique to deploy stochastic policy gradient methods where the stochastic policy is supported by a stochastically disturbed constrained parametric optimization problem. This approach allows one to restrict the support of the stochastic policy to a safe set described via constraints. In particular, robust Nonlinear Model Predictive Control, where safety requirements can be imposed explicitly, can be selected as a parametric optimization problem. Imposing restrictions on the support of the stochastic policy creates some technical challenges when computing the gradient of the policy score function required in the computation of the stochastic policy gradient. Computationally inexpensive methods are proposed here to tackle these challenges, using interior-point methods and techniques from parametric Nonlinear Programming. The specific case of robust linear Model Predictive Control, where the prediction model is linear, is further developed, and a methodology to impose safety requirements throughout the learning is proposed. The proposed techniques are illustrated in simple simulations, showing their behavior. This paper has a companion paper \cite{Gros2019b} investigating the deterministic policy gradient approach in the same context as in this paper. The stochastic policy gradient approach is theoretically simple and appealing, and requires less assumptions that its deterministic counterpart presented in \cite{Gros2019b}. It also requires solving a single NLP per time instant, as opposed to two in the deterministic case. However, the computational complexity of the stochastic approach presented here can be significantly higher that the deterministic policy approach of \cite{Gros2019b} when the number of parameters $\vect\theta$ is larger than the input size $n_{\vect a}$. This effect is a direct result of the computational complexity of evaluating the second-order sensitivities required to form the gradient of the stochastic policy score function, see \eqref{eq:SecondOrderSens:Stochastic}. 

%\end{IEEEproof}
\bibliographystyle{plain}
\bibliography{syscop}

\begin{IEEEbiography}[{\includegraphics[width=1in,height=1.25in,keepaspectratio]{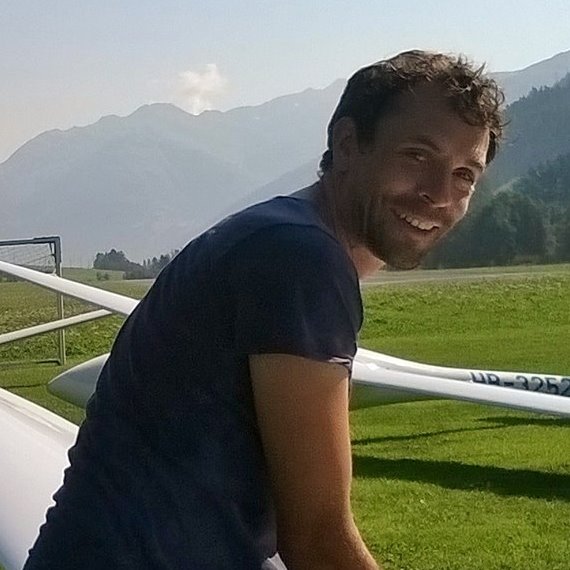}}]{S\'ebastien Gros}
	received his Ph.D degree from EPFL, Switzerland, in 2007. After a journey by bicycle from Switzerland to the Everest base camp in full autonomy, he joined a R\&D group hosted at Strathclyde University focusing on wind turbine control. In 2011, he joined the university of KU Leuven, where his main research focus was on optimal control and fast NMPC for complex mechanical systems. He joined the Department of Signals and Systems at Chalmers University of Technology, G\"{o}teborg in 2013, where he became associate Prof. in 2017. He is now full Prof. at NTNU, Norway and guest Prof. at Chalmers. His main research interests include numerical methods, real-time optimal control, reinforcement learning, and energy-related applications.
\end{IEEEbiography}

\begin{IEEEbiography}[{\includegraphics[width=1in,height=1.25in,clip,keepaspectratio]{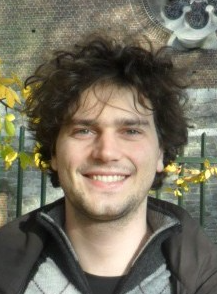}}]{Mario Zanon}
	received the Master's degree in Mechatronics from the University of Trento, and the Dipl\^{o}me d'Ing\'{e}nieur from the Ecole Centrale Paris, in 2010. After research stays at the KU Leuven, University of Bayreuth, Chalmers University, and the University of Freiburg he received the Ph.D. degree in Electrical Engineering from the KU Leuven in November 2015. He held a Post-Doc researcher position at Chalmers University until the end of 2017 and is now Assistant Professor at the IMT School for Advanced Studies Lucca. His research interests include numerical methods for optimization, economic MPC, optimal control and estimation of nonlinear dynamic systems, in particular for aerospace and automotive applications.
\end{IEEEbiography}

\end{document}